%% file: MatchingNetworkvExtendedRevised2.tex
\documentclass[lettersize,journal]{IEEEtran}

\input epsf
\usepackage{graphicx}
\usepackage{url}
\usepackage{multirow}
\usepackage{array}
\usepackage{import}
\usepackage{pgf}
\usepackage{xcolor}
\usepackage{upgreek}
\usepackage{background}
\usepackage[final]{changes}
\usepackage{cancel}
\usepackage{amsmath}

\SetBgContents{Conference ID 56477X}
\SetBgScale{1}
\SetBgAngle{0}
\SetBgColor{black}
\SetBgOpacity{1}
\SetBgPosition{current page.north east}
\SetBgHshift{-2cm}
\SetBgVshift{-1cm}

\IEEEpubid{\fontfamily{phv}\selectfont \small \begin{tabular}[t]{@{}l@{}}\copyright 2024 IEEE Personal use of this material is permitted. Permission from IEEE must be obtained for all other uses, in any current or future media,\\
    including reprinting/republishing this material for advertising or promotional purposes, creating new collective works,  \\ for resale or redistribution to servers or lists, or reuse of any copyrighted component of this work in other works. \\ Hoffmann, Johannes, et al. "Comparison of Impedance Matching Networks for Scanning Microwave Microscopy." \\ IEEE Transactions on Instrumentation and Measurement (2024) DOI 10.1109/TIM.2024.3378310 \end{tabular}}
\newcommand{\STAB}[1]{\begin{tabular}{@{}c@{}}#1\end{tabular}}

\begin{document}

\title{\LARGE Comparison of Impedance Matching Networks for Scanning Microwave Microscopy}
\bibliographystyle{IEEEtran}



%
\author{\IEEEauthorblockN{Johannes Hoffmann, Sophie de Préville, Bruno Eckmann, Hung-Ju Lin, Benedikt Herzog,
Kamel Haddadi, Didier Théron,  Georg Gramse, Damien Richert, José Morán-Meza and
François Piquemal} \\
}



\maketitle

\begin{abstract}
In this paper, a definition of the gain and added noise of impedance matching networks for scanning microwave microscopy is given. This definition can be used to compare different impedance matching techniques independently of the instrument used to measure the S-parameter. As a demonstration, impedance matching devices consisting of a Beatty line, a tuner, and interferometric setups with and without amplifiers have been investigated. Measurement frequencies up to 28\,GHz are used, and the maximal resulting gain found was 9504.7 per Siemens.
\end{abstract}

%
\IEEEpeerreviewmaketitle

\pagenumbering{gobble}

\section{Introduction}
In scanning microwave microscopy (SMM), a vector network analyzer (VNA) is connected through an impedance-matching network to a modified atomic force microscope (AFM) supporting a conductive tip. The VNA sends a signal through the matching network to the AFM tip. There, the signal is reflected back to the VNA as a function of the material properties or local impedance below the tip. Microwave hardware is usually built for 50\,$\Omega$ characteristic impedance, but the impedances to be measured at the tip are in the range of  k$\Omega$. This impedance mismatch needs to be compensated for by a matching network to enable low noise and high signal measurements.

In one of the pioneering publications in the area of scanning microwave microscopy, \cite{tabib1999evanescent}, the impedance matching is done with a microstrip resonator.  In the following decades, various methods for creating such matching networks have been researched.  In \cite{kim2002development}, a tunable coaxial $\lambda/4$ resonator is used as a matching circuit and its influence on electrical sensitivity and spatial resolution are investigated. In \cite{randus2010novel} and \cite{haase2023uncertainties}, general techniques of measuring very high impedances with a VNA using different couplers are discussed, and comparisons of measurement noise with and without the matching network are made. In \cite{bakli2014interferometric}, an SMM with a power divider based interferometer and, in \cite{gu2016setting}, a hybrid coupler based interferometric setup are investigated for operating parameters and data quality. In \cite{tuca2017interferometer}, \cite{horibe2017demonstrations}, and \cite{horibe2018measurement}, comparisons between no matching network, a network consisting of a shunt and a $\lambda/2$ line, and a 90 degree hybrid in combination with a low noise amplifier have been made. \added{The setups described in \cite{tabib1999evanescent} - \cite{horibe2018measurement} contain  a low noise amplifier (LNA) either inside a vector network analyzer (VNA) or external as the first amplification stage. In none of the cited references the noise characteristics of the first amplification stage is specified.  According to the Friis formula, \cite{meinke2013taschenbuch},  the first amplification stage largely determines noise behavior. Thus, it is very difficult to compare the results without known noise characteristics of the first amplification stage. E.g., a very good matching network followed by a bad amplification stage can result in bad performance and vice versa.}   In the proceedings paper \cite{hoffmann2022Impedance}, the authors of the present paper  described a first measurement technique for comparable characterization of matching networks in SMM. \added{Here, the initial proceedings paper has been extended by adding three additional measurement setups and results, including retraction measurements, and additional theory on the computation of expectable gains.}

\added{What the studies on improving existing matching networks have in common is that the results of different studies are not comparable because the noise of the VNA or of the first amplification stage was not taken into account.} Here, matching networks consisting of a Beatty line, a tuner, and two interferometric setups have been investigated. A theoretical calculation of the expected gain is presented. A technique to practically quantify the gain and added noise of a given matching network is introduced. These figures of merit enable comparison and application dependent selection of network topologies.  Additionally, two different techniques to obtain the raw data are tested. One works with a retraction scan where different capacitance values are measured at different heights. The second is based on a calibration sample containing known capacitors.\IEEEpubidadjcol

  \begin{figure}
    \center
    \fontsize{7pt}{7pt}
     \def\svgwidth{3in}
     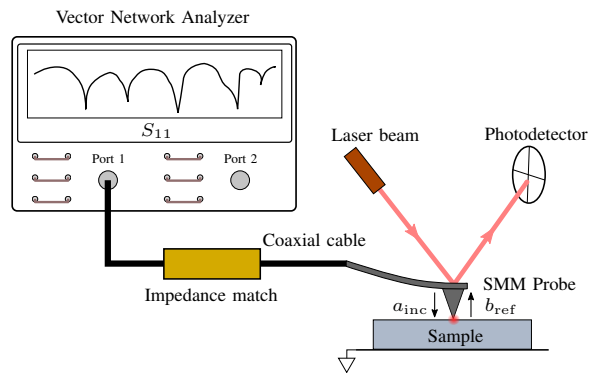
     \caption{\small Schematic of SMM setup at METAS and JKU. In the case of METAS the matching element between VNA and tip is a Beatty line, a 25\,$\Omega$ mismatch section. In the case of JKU the matching element is a tuner. }
    \label{figMJSetups}
     \end{figure}

The authors use the following impedance matching networks in their labs and test these setups with the aforementioned technique. The setup at the \added{Federal Institute of Metrology METAS (METAS)} consists of a Rohde Schwarz ZNA 50 with a Beatty line (Maury 7942C25) connected to a Nanosurf FlexAFM with an SMM cantilever; see Fig. \ref{figMJSetups}. The Beatty line is in essence a piece of 25\,$\Omega$ transmission line which produces reflections that, at certain frequencies, minimize the reflection coming from the tip. In terms of a matching network, this is most probably the approach requiring the fewest parts and thus the most stable over time. 

The setup at the \added{Johannes Keppler University (JKU)} consists of a 50\,GHz Keysight PNA L with frequency extension and a Keysight SMM. The impedance matching network consists of a Mauri 1643P tuner; see Fig. \ref{figMJSetups}. The tuner can improve the match at arbitrary frequencies. This setup does not contain active elements apart from the VNA and is expected to be stable over time.

\begin{figure}
  \center
  \fontsize{7pt}{7pt}
   \def\svgwidth{3in}
   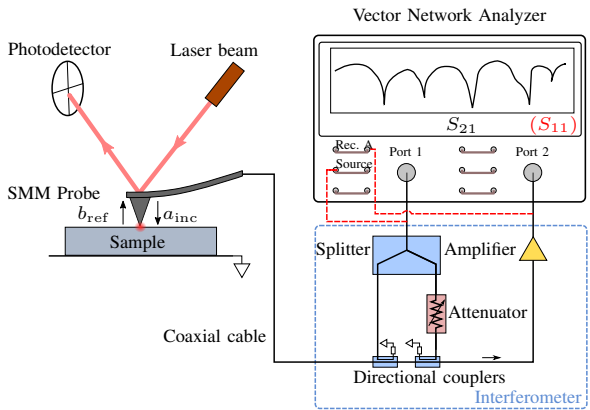
   \caption{\small  Schematic of SMM setup at LNE and the University of Lille. In the case of LNE the interferometer is directly connected to the test ports whereas the University of Lille uses direct receiver access. }
   \label{figLLSetups}
   \end{figure}

\begin{figure}
 \center
  \includegraphics[width=3in]{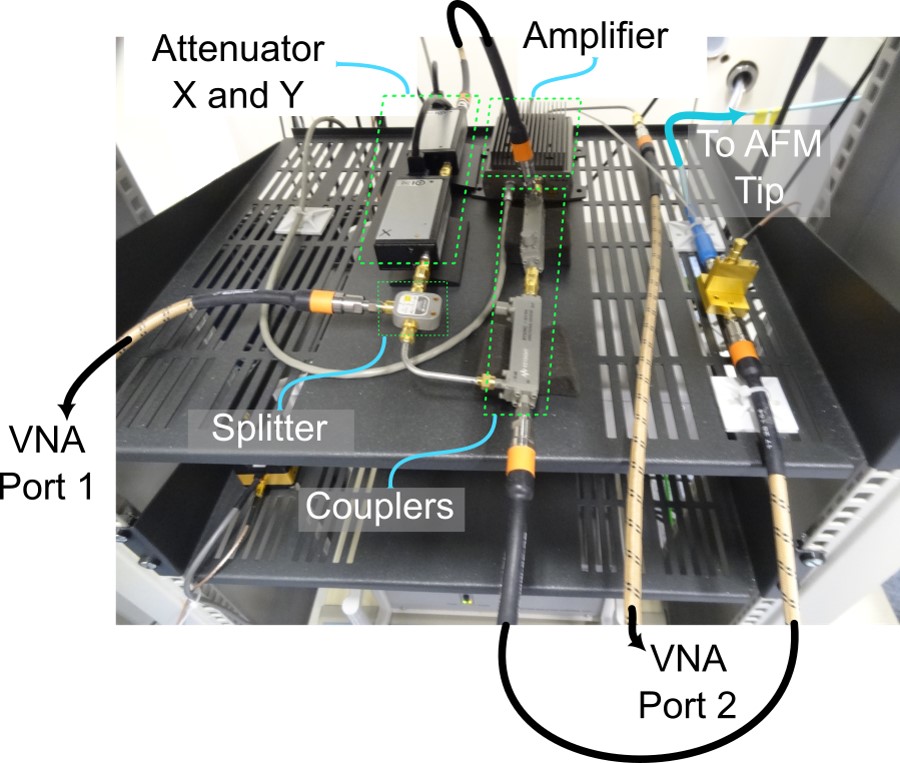}
  \caption{\small Photograph of the Mach-Zehnder-based interferometric impedance matching system at LNE.}
  \label{figLNEInterf}
  \end{figure}

The matching network at \added{Laboratoire national de métrologie et d'essais (LNE)} is a home-built Mach-Zehnder-based interferometric system (see Figs. \ref{figLLSetups} and \ref{figLNEInterf}) using  microwave electronic devices purchased from Keysight Technologies. The interferometer was built using a coaxial divider (model 11636B: DC – 26.5\,GHz, power divider), two coaxial directional couplers (model 87300C: 1 – 26.5\,GHz), two programmable step attenuators (model 84904K: DC – 26.5\,GHz, 22\,dB attenuation total range, selected attenuation=7\,dB and 21\, dB), and a microwave system amplifier (model 83017A: 0 – 26.5\,GHz, 25\,dB minimum gain). The VNA is a P9374A.

\begin{figure}
  \center
  \fontsize{7pt}{7pt}
   \def\svgwidth{3in}
   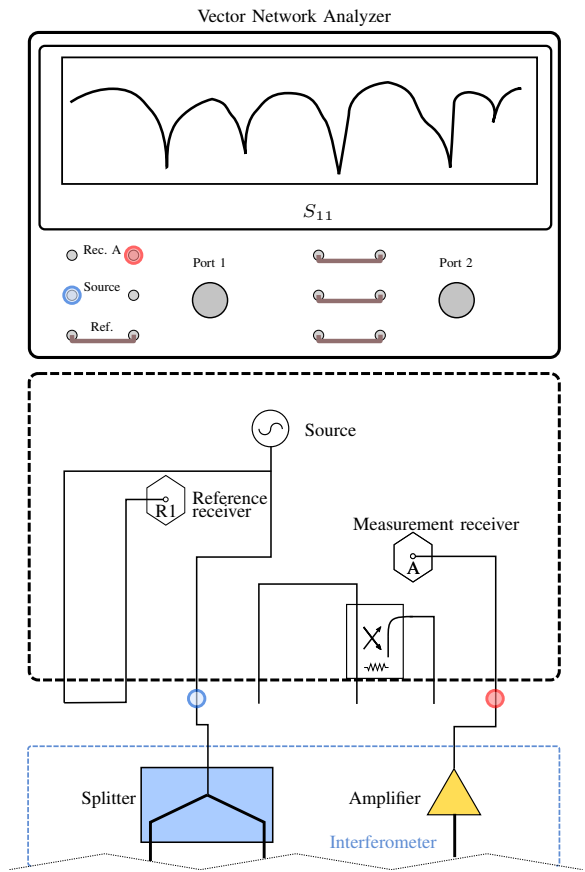
   \caption{\small Illustration of direct receiver access at port one. Instead of connecting the divider to the test port, it is directly connected to the source (\added{blue circle}) and the amplifier is directly connected to the receiver (\added{red circle}) at port one.}
   \label{figDirectREC}
   \end{figure}

The setup at the University of Lille is schematically the same, see Fig. \ref{figLLSetups}, as the one used at LNE but with direct receiver access and a different VNA; see Fig. \ref{figDirectREC}. The advantage of this configuration is that only one port of the VNA is occupied, and the \added{signal-to-noise ratio (SNR)} is better because coupling losses at the test port coupler are avoided. These coupling losses at the test port without direct receiver access are approximately 13\,dB in the VNA of the University of Lille. \added{The matching network at  the University of Lille is a home-built Mach-Zehnder-based interferometric system using commercial and non-commercial  microwave electronic devices. The interferometer was built using a coaxial splitter from Narda (unknown model number), two coaxial directional couplers (Narda 10 dB, 2-18\,GHz), a programmable PIN diode as attenuator (PIN diode provided by Keysight attenuation between 0 and 5 dB), and a microwave system amplifier (MicroSemi AML218L2502 with gain of approximately 30\,dB plus, a preamplifier of 12\,dB from Agilent (no reference, internal product). The VNA is a Keysight E8363B 10\,MHz - 40\,GHz.}

\section{Theory and Techniques}
The objective of the described techniques is to extract the gain of the matching network and its added noise. \added{In the following, the used techniques and the measurement uncertainties associated to these techniques are described. } 
\subsection{Gain Measurement}
In order to determine the gain, one measures with the SMM two different admittances. This can be done in a retraction scan or while scanning over a sample with known admittances. \added{While this approach is relatively simple for the sample with known admittances this is more complicated for the retraction scan. During the retraction scan, a signal with amplitude $V_0$ and a low frequency $\omega$ in the kHz range is applied to the tip and sets the tip into motion. The motion can be measured with a lock-in amplifier; see also \cite{gramse2014calibrated}.  The relation
between the electrostatic force, $F_{es}$, and the tip–sample capacitance, $C$ is
\begin{equation}
  F_{es} =\frac{1}{4}\frac{dC}{dz}V_0^2cos(2\omega t).
\end{equation}
The electrostatic force can be converted into the voltage-normalized force 
\begin{equation}
  \frac{dC}{dz} = \frac{4 F_{es,2\omega}}{V_0^2} \label{equdcdz}
\end{equation}
that corresponds to the capacitance gradient with respect to the tip–sample distance, $z$. \added{Note that equation (\ref{equdcdz}) is from \cite{gramse2014calibrated}. In \cite{gramse2014calibrated} it contains an error which is corrected here.}  Integrating this curve using a standard numerical integration technique (trapezoidal rule),  yields the desired tip–sample capacitance, $C(z)$, and using (\ref{equadmtip}) the  admittance, $Y{(z)}$, which can be used to determine the gain. From this retraction scan or image,} one can extract the raw reflection $S_{11}$ or transmission $S_{21}$ of two different known admittances at the tip. The gain of the  matching network can  now be \added{defined} as
\begin{equation}
  G=\frac{|S_1-S_2|}{|Y_1-Y_2|}.\label{eqgain}
\end{equation}
Here, $Y_n$ represents the two different measured admittances on the sample or during the retraction scan, whereas $S_n$ stands for the S-parameter measured at the respective standard or during a retraction scan. This assumes a calibrated VNA.  Often, it is enough to use the factory settings because they let the VNA appear to be roughly calibrated.
\subsection{Gain of Passive Matching Networks}
In this paper, two types of matching networks are investigated. LNE and Lille use interferometry based matching networks, whereas METAS and JKU use matching networks which are put in between tip and VNA. For this latter type, assume that the capacity at the tip is $C_{tip}$ and the operating frequency is $f$. Then the admittance of the tip capacity is
\begin{equation}
  Y=j 2 \pi f C_{tip}. \label{equadmtip}
\end{equation}
The reflection coefficient of this admittance is 
\begin{equation}
  \Gamma=\frac{1/Y-50\Omega}{1/Y+50 \Omega} \label{equadm}.
\end{equation}
\added{The capacitances typically measured with an SMM are between 0.1\,aF and 10\,fF which translates approximately to admittances between 0.6\,nS up to 3\,mS assuming frequencies between 1\,GHz and 50\,GHz. Retraction scans yield admittances in the range of nano to micro Siemens, whereas measurements on substrates typically yield admittances in the micro to milli Siemens range. Thus one can assume the admittance to be small and we approximate (\ref{equadm}) as} 
\begin{equation}
  \Gamma \approx 1-100\Omega * Y \label{eqgamma}.
\end{equation}
Note that this approximation yields a reflection coefficient magnitude larger than one. While this is not physically possible, it poses no problem for the following calculations. Using (\ref{eqgamma}) with (\ref{eqgain}), one can show that the gain of a lossless system without a matching network would be 100 per Siemens. \added{Values for lossy systems can be below 50 per Siemens.} 

Now a matching network is inserted between the tip and the VNA. \added{The primary purpose of the matching network is to match the high impedance of the tip to an impedance close to 50\,$\Omega$. This is necessary because at this impedance many VNAs and LNAs have their optimal noise performance. A secondary purpose of the matching network is to amplify the signal.} The matching network can be described by its four S-parameters $S_{xyM}$ \added{and the match (input reflection factor) of the VNA or low noise amplifier $\mu$. Using Mason's rules, this} converts the reflection coefficient to
\begin{eqnarray}
  \Gamma_M &=\frac{S_{11M}(1-S_{22M}\Gamma)+S_{21M}S_{12M}\Gamma}{\Delta}. \label{eqIMMatch} \\
  \Delta &=1-S_{22M}\Gamma-S_{11M}\mu \ldots \nonumber\\
  &-S_{21M}S_{12M}\Gamma\mu  + S_{11M} \mu S_{22M}\Gamma. \nonumber
\end{eqnarray}
\added{As already said, most often the optimal input reflection coefficient for low noise of LNAs or VNAs is close to zero. Thus (\ref{eqIMMatch}) should be set close to zero by choosing the $S_{xyM}$ accordingly. Note that for lossy networks the following holds: $|S_{xx}|<\sqrt{1-|S_{yx}|^2}$. Under these constraints, $\Gamma_M=0$ and constrained amplitudes of transmission and reflection, a third goal should be persued. This is that $\Gamma_M$ depends as strongly as possible on $\Gamma$, i.e., that the gain is high. In practice, these objectives are pursued by building or simulating matching networks that contain one or several line elements with reflections on both sides of each line element. Then a frequency sweep is made and the gain and amplitude of $\Gamma_M$ over frequency are recorded. Finally, frequencies with favorable gain and noise characteristics are used for operation. A closer look at (\ref{eqIMMatch}) shows that the gain is not linear because $\Gamma$ appears in the denominator. The nonlinearity of the gain can be found from simulating (\ref{eqIMMatch}) and is in practice below 10\,\% for the setups discussed in this paper when regarding a scale from 0\,mS to 3\,mS. The nonlinearity can be higher for other setups.} 
\subsection{Gain of Active Matching Networks}
Interferometric matching networks require a different gain calculation. For simplicity, one assumes that all components do not have reflections. The measured signal at the VNA \added{can then be deduced from Fig. \ref{figLLSetups} as}
\begin{equation}
  \Gamma_{MI}=(A_{divider}A_{coupler}\Gamma+CS)A_{amp}. \label{eqinteferometric}
\end{equation}
Here $A_{divider}$, $A_{coupler}$, $CS$, and  $A_{amp}$ denominate divider attenuation, coupler attenuation, cancellation signal, and amplification, respectively. The cancellation signal is usually set such that the input of the amplifier is close to zero. The quantities in (\ref{eqinteferometric}) are obtained from the components respective data sheets: $A_{divider}=0.447$ $A_{coupler}=0.32$, and $A_{amp}=56.23$. Note that the amplifier has 25\,dB minimum amplification with $\pm$5\,dB flatness. Inserting these numbers into (\ref{eqinteferometric}) and plugging this into (\ref{eqgain}) yields a gain of 804 per Siemens. Reflections, resonances, and unaccounted losses between the components can lower or increase this gain. 
\subsection{Noise Measurement}
In order to characterize the noise, one would lift the tip high above the sample and do a scan in air. This yields the noise of the complete electrical system, including VNA and matching network. Note that the translation of mechanical noise to electrical noise is strongly attenuated by lifting the tip. Consequently this part is missing in what is called here noise of the complete setup $N_{\rm{C}}$. By calculating the variance and standard deviation of the data, one gets the noise 
\begin{equation}
  N_{\rm{C}}=\sqrt{var{(S)}}. \label{eqNC}
\end{equation}
Here $S$ stands for the raw S-parameters in each pixel of the image scanned in air. Another way to obtain the noise of the complete system would be to fit a polynomial of higher order (here, order 5) into the raw S-parameters of a retraction curve and thus extract the noise of the VNA and matching network by taking the difference between the polynomial and measured data. Again here the noise does not contain the part which translated from mechanical noise to electrical noise.

\begin{figure}
  \center
  \resizebox{3.2in}{!}{\input{abs_s11.pgf}}
\resizebox{3.2in}{!}{\input{phase_s11.pgf}}
  \caption{\small Measurement results of a sample containing micro disc capacitors from MC2 Technologies. Upper and lower panel show raw results in amplitude and phase in radian. The intermediate frequency bandwidth is set to 500\,Hz. The raw results have been flattened with a second order polynomial and a plane has been subtracted.}
  \label{figMC2Sample}
  \end{figure}
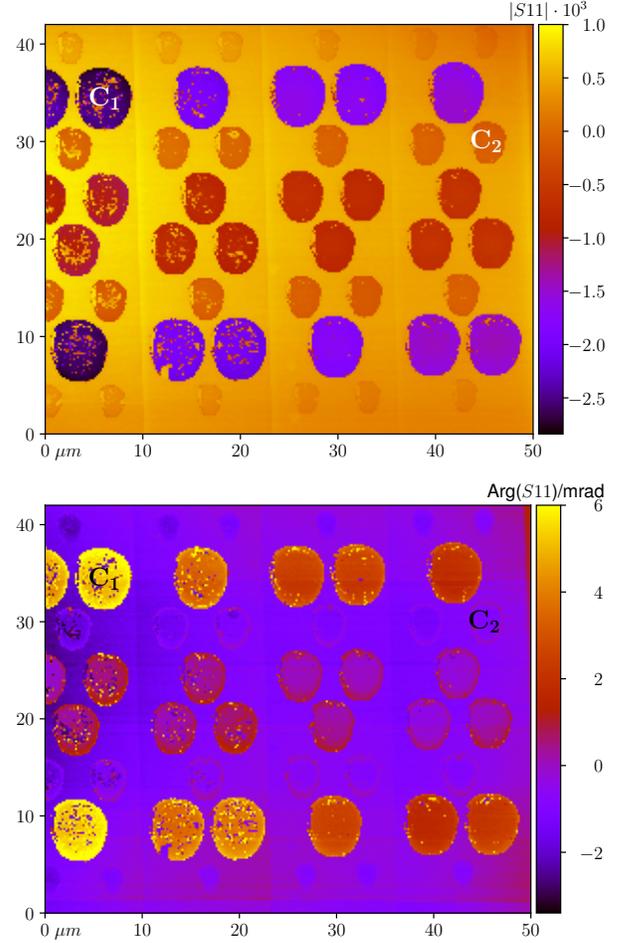

In the last step, one needs to characterize the noise of the VNA. This can be done as proposed in VNA Tools framework, \cite{wollensack2013metas}, or by connecting short circuits to the VNA ports, and measuring the noise in $S_{21}$ data, which yields the noise floor $N_{\rm{F}}$, and the noise in $S_{11}$, which yields the trace noise in amplitude $N_{\rm{TA}}$ and phase $N_{\rm{TP}}$
\begin{eqnarray}
  \label{eq3-5}
  N_{\rm{F}} &= &\sqrt{var{(S_{21})}} \label{equFloor}\\
  N_{\rm{TA}} &=& \sqrt{var{(|S_{11}|)}} \label{equtracea}\\
  N_{\rm{TP}} &=&\sqrt{var{(arg{(S_{11})})}}.\label{equtracep}
\end{eqnarray}
\added{This approach is also proposed in \cite{VNAGuide} but with the difference that (\ref{equFloor}) looks at the distance between data points and their mean whereas in \cite{VNAGuide} the noise floor is decomposed into real and imaginary parts. Equations (\ref{equtracea}) and (\ref{equtracep}) assume the reflection amplitude of the used short circuits to be roughly one. To find the noise of the VNA one needs to add the distributions representing noise floor and trace noise.} The noise contribution of the VNA per measurement is
\begin{equation}
\label{eq_N_VNA}
N_{\rm{VNA}}=\sqrt{N_{\rm{F}}^2+N_{\rm{TA}}^2|S_m|^2+|S_m|^2 N_{\rm{TP}}^2} \\
\end{equation}
where the mean of the $S$ used in (\ref{eqNC}) defines \mbox{$S_m=\bar{S}$} \added{and noise floor and trace noise are assumed to be uncorrelated}.
The noise of the VNA needs to be subtracted from the complete noise $N_{\rm{C}}$ in order to obtain the noise of the matching network
\begin{equation}
\label{eq_N_M}
N_{\rm{M}}=\sqrt{N_{\rm{C}}^2-N_{\rm{VNA}}^2},
\end{equation}
which is always real because $N_{\rm{C}}>N_{\rm{VNA}}$. Now each matching network can be characterized by its gain $G$ and by its added noise $N_{\rm{M}}$.

\section{Results}
\subsection{METAS result}
At METAS, this technique has been applied to an SMM with a Beatty line as a matching network attached to it. \added{The sample described in \cite{piquemal2021progress} contains micro-capacitors and was used to determine the gain factor.}  In Fig. \ref{figMC2Sample}, one can see the absolute value of raw $S_{11}$ and phase measured at 2.158\,GHz. Similar results have been obtained at 3.62\,GHz and 5.395\,GHz. The capacitor $C_1$ in Fig. \ref{figMC2Sample} with an assumed radius of 2\,$\upmu$m and a silicon oxide height of 50\,nm and the capacitor $C_2$  with a radius 1\,$\upmu$m and an oxide height of 200\,nm have been used as references. The top left capacitor has a capacitance of $C_1=9.52$\,fF and the right third row capacitor has $C_2=0.86\,$fF assuming a relative permittivity of $\epsilon_{\rm{SiO2}}=3.9$ and using the equations given in \cite{sloggett1986fringing}. \added{Please note that measurement results from very small capacitors are not used because they are subject to slightly higher measurement uncertainties. The combined measurement uncertainty for the capacitors is in the order of 3\,\%, see \cite{piquemal2021progress}. This translates to an uncertainty component in gain of 4.5\,\% and combined with the uncertainty from linearity (10\,\%) yields a total uncertainty of 11\,\% for the measured gain.}  Evaluating the image at the positions of the chosen capacitors yields the  gain per Siemens given in Table \ref{tabGain}. \added{The frequencies, that have the highest gain within the available frequency range of the instrument, are used for characterization. Note that the frequencies in between these listed frequencies have usually much worse characteristics.}

\begin{table}[h]
	\caption{Gain per S of  SMM setups at best frequencies}
  \label{tabGain}
  \begin{center}
		\begin{tabular}{c | c c c }
      \multicolumn{4}{c}{METAS} \\ \hline
      Frequency in GHz	& 2.158 & 3.62 & 5.395	\\
      Gain per S	 & \added{44} &  \added{111} &  \added{261} \\ \hline
      \multicolumn{4}{c}{JKU} \\ \hline
      Frequency in GHz	& 3  & 8 & 28 \\
      Gain per S	 & 105 &  177 &  242 \\ \hline
			  \multicolumn{4}{c}{LNE} \\
        \hline
			 		Frequency in GHz 	& 2.13 &  4.54	& 5.42\\
			Gain per S	 & \added{310}  &   \added{737} & \added{897} \\
      	 		Frequency in GHz 	& 13.81 & 15.89 & 18.94 \\
			Gain per S	  & \added{4019}  & \added{9505}  &  \added{7907} \\ \hline
      \multicolumn{4}{c}{Lille} \\
      \hline
      Frequency in GHz 	& 2.16 & 4.465  &  \\
Gain per S	  & 1030  & 620  &  \\
		\end{tabular}
  \end{center}
\end{table}

\begin{figure}
  \center
  \includegraphics[width=3in]{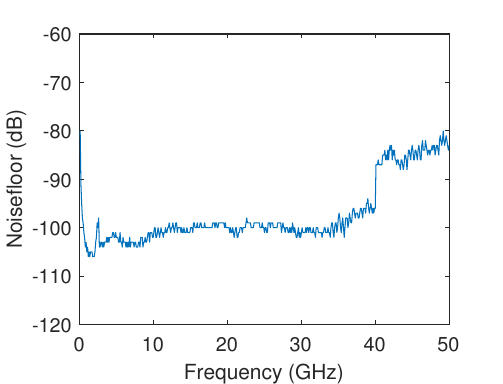}
  \caption{\small Noise floor in dB of a Rohde Schwarz ZNA 50 with 500\,Hz intermediate frequency bandwidth.}
  \label{figNoisefloor}
  \end{figure}

The noise floor and trace noise of the VNA are measured, and the noise of the complete system consisting of VNA and matching network is measured in order to determine the amount of added noise by the matching network. The results of complete noise $N_{\rm{C}}$, VNA noise $N_{\rm{VNA}}$, and added noise $N_{\rm{M}}$ are given in Table \ref{tabNoise}. \added{The VNA noise has been determined from 801 points and thus the expected standard deviation of this noise measurement is 0.3\,dB. This is calculated using $\Delta \sigma=\sigma / \sqrt{2n-2}$ where $\sigma$ is the standard deviation of the to-be-estimated distribution and $\Delta\sigma$ is the standard deviation of this estimation due to using $n$ points of the distribution. The complete noise has been extracted from an image with 256x256 pixels. This leads to a standard uncertainty of 0.03\,dB in this measurement. For simplicity, all noise quantities are rounded to 1\,dB.}  Measuring the noise floor of the VNA yielded the results depicted in Fig. \ref{figNoisefloor}. 

\begin{table}[h]
	\caption{Noise of SMM setups at best frequencies}
  \label{tabNoise}
	\begin{center}
		\begin{tabular}{c c| r r r }
      & Frequency in GHz	& 2.158 & 3.62 & 5.395	\\ \hline
      \multirow{3}{*}{\STAB{\rotatebox[origin=c]{90}{METAS}}}
   &\added{$N_{\rm{C}}$	in dB} & \added{-86}  &  \added{-82}  &  \added{-78} \\
   &\added{$N_{\rm{VNA}}$ in dB} & \added{-103} & \added{-102} & \added{-98}\\
   &\added{$N_{\rm{M}}$ in dB} & \added{-86} & \added{-82} & \added{-78} \\ \hline
   &Frequency in GHz	& 3  & 8  & 28 	\\   \hline
   \multirow{3}{*}{\STAB{\rotatebox[origin=c]{90}{JKU}}}
&\added{ $N_{\rm{C}}$ in dB} & \added{-89}  &  \added{-95}  &  \added{-79} \\ 
&\added{ $N_{\rm{VNA}}$ in dB	} & \added{-101}  &  \added{-103}  &  \added{-92} \\ 
&\added{ $N_{\rm{M}}$ in dB}	 & \added{-90}  &  \added{-96}  &  \added{-80} \\ \hline
			& Frequency in GHz	& 2.13 &  4.54	& 5.42\\	\hline
      \multirow{3}{*}{\STAB{\rotatebox[origin=c]{90}{LNE}}}
			&\added{ $N_{\rm{C}}$ in dB}  & \added{-65}  &    \added{-60} & \added{-64} \\
			&\added{ $N_{\rm{VNA}}$ in dB} & \added{-95} &  \added{-94} & \added{-95}\\
			&\added{ $N_{\rm{M}}$ in dB} & \added{-65} & \added{-60} & \added{-64}\\	 \hline 
			& Frequency in GHz	& 13.81 & 15.89 & 18.94 \\ 	\hline
      \multirow{3}{*}{\STAB{\rotatebox[origin=c]{90}{LNE}}}
			&\added{$N_{\rm{C}}$ in dB}  & \added{-65}  & \added{-73}  &  \added{-68} \\
			&\added{$N_{\rm{VNA}}$ in dB} & \added{-96} & \added{-96} & \added{-96} \\
			&\added{$N_{\rm{M}}$ in dB }& \added{-65} & \added{-73} & \added{-68} \\ \hline
      & Frequency in GHz	& 2.16 & 4.465 &   \\ 	\hline
      \multirow{3}{*}{\STAB{\rotatebox[origin=c]{90}{Lille}}}
			&\added{$N_{\rm{C}}$ in dB }	 & \added{-43} & \added{-50}  &   \\
			&\added{$N_{\rm{VNA}}$ in dB } & \added{-95} &\added{-97} &  \\
			&\added{$N_{\rm{M}}$  in dB  } & \added{-43} & \added{-50} &  
		\end{tabular}
	\end{center}
\end{table}

\subsection{JKU result}
\begin{figure}
  \center
  \includegraphics[width=3in]{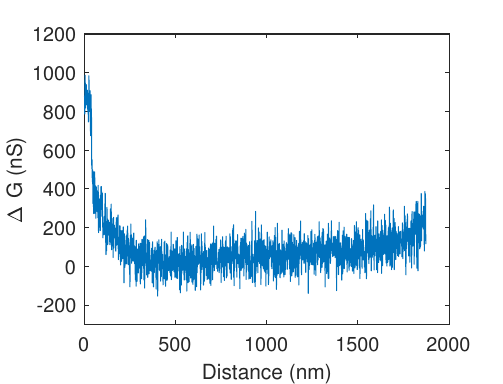}
  \caption{\small Offsetted conductance $\Delta G$  plotted over the distance between sample and tip at 3\,GHz with a tuner as impedance match attached. \added{$\Delta G$ is measured with the JKU setup.}}
  \label{figdeltag}
  \end{figure}

\begin{figure}
    \center
    \includegraphics[width=3in]{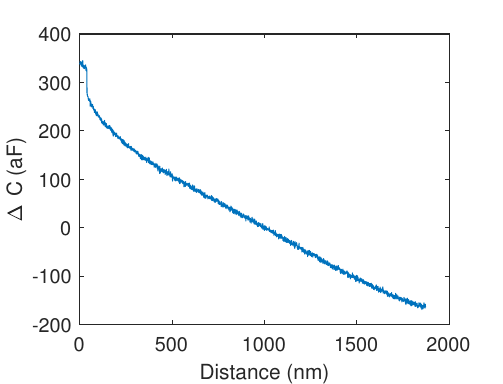}
    \caption{\small Offsetted capacitance $\Delta C$  plotted over the distance between sample and tip at 3\,GHz with a tuner as impedance match attached. \added{$\Delta C$ is measured with the JKU setup.}}
    \label{figdeltac}
    \end{figure}

At JKU, combined retraction scans of electric force microscopy (EFM) and SMM of the metallic tip over a \added{highly doped silicon} surface have been executed. \added{The EFM signal is used to extract calibrated values for the tip-sample capacitance, see (\ref{equdcdz}).  One calculates the associated \added{admittances $Y(z)$}, from these capacitance values.} \added{The SMM is calibrated}, using the \added{modified short open load calibration technique (mSOL)} and \added{admittances $Y(z)$} as standards. \added{This results in} calibrated retraction curves in $S_{11}$. \added{Using (\ref{equadm}), these are converted} into calibrated retraction curves in conductance  $G$ and capacitance $C$; see Figs. \ref{figdeltag} and \ref{figdeltac} \added {for JKU results}. \added{When approaching the tip towards the surface one observes mainly a change in the capacitance channel and the conductance channel stays constant. Only when the tip jumps into contact one sees a slight change in the conductance channel that is associated with loss of the microwave signal in the doped silicon. From Figs.  \ref{figdeltag} and \ref{figdeltac} signal changes are extracted and the gain is calculated by taking the signal differences between the z-positions of approximately 1\,nm and 1800 nm. The extracted gain of the setup with the tuner at 3\,GHz is given in Table \ref{tabGain}.}

The complete noise $N_C$ is calculated from fitting a fifth order polynomial \added{using an unweighted least squares technique, \cite{reference.wolfram_2023_nonlinearmodelfit},} to the raw S-parameters of the retraction scan\added{, and applying (\ref{eqNC}) to the residuals of this fitting}. \added{Note that the polynomial is only fitted to $z$-distances larger than 100\,nm, see as well Figs. \ref{figdeltag} and \ref{figdeltac}. The complete noise has been extracted from a retraction curve with 2000 points and thus the standard deviation on this measurement is 0.2\,dB.} The noise of the VNA alone is measured with VNA Tools. \added{The VNA noise has been measured with 801 points which leads to a standard deviation of 0.3\,dB.} These two noise measurements are used to calculate the noise of the matching network, see Table \ref{tabNoise}. \added{For simplicity, all values have been rounded to 1\,dB.} For comparison, measurements at two frequencies (8\,GHz and 28\,GHz) of the same setup without the tuner are given.

\subsection{LNE and University of Lille result}
\begin{figure}[h]
	\center
	\includegraphics[width=3in]{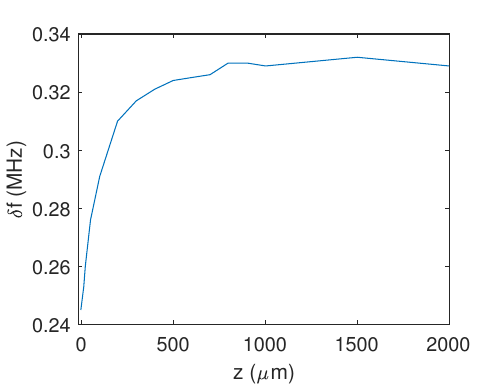}
	\caption{\small \added{Shift of resonance peak frequency $\delta f$} of $S_{11}$ parameter as a function of tip-sample distance $z$.}
	\label{fig:f0_function_of_tipsample_distance}
\end{figure}

At LNE and the University of Lille, the noise of the complete setup, free from the sample's influence, is measured before estimating the added noise. To this end LNE recorded the $\ S_{11}$ parameter at several tip-sample separation distances and determined the corresponding values of the resonance frequency $f_0$. Fig. \ref{fig:f0_function_of_tipsample_distance} shows that $f_0$ increases with the tip-sample distance and reaches a plateau at a distance of 1000$\,\mu$m. No further changes in $\ f_0$ were observed as the distance increased. Therefore, the tip-sample distance of 1000$\,\mu$m was considered the critical separation distance above which the influence of the sample on the measured $S_{11}$ parameter vanishes. 
Consequently, at LNE and the University of Lille the complete electrical noise $N_C$ for each scan frequency is acquired by placing the probe at a tip-sample separation distance greater than 1000$\,\mu$m \added{and taking one image. The $N_C$ is determined using (\ref{eqNC}).} Furthermore, the noise floor and the trace noise  are recorded using the VNA Tools framework with two short circuit standards (HP85052-60006 and MMC8047F F791). At LNE the equivalence of the short standards for noise characterization has been verified by exchanging the standards between port one and two. Finally, the noise associated with the setups is computed using equations (\ref{eqNC}-\ref{eq_N_M}). The results are shown in Table \ref{tabNoise}. \added{The results from LNE and the University of Lille have the same expected standard uncertainty as for METAS because the number of points and pixels for determining the noise contributions were the same.} 

As LNE and the University of Lille use interferometric systems in their setups, the output signal is amplified, and therefore there is the possibility to saturate the receivers of the VNA. To stay below the saturation region, LNE uses an attenuator in their setup. SMM scans were acquired at different frequencies on an MC2 sample, and the attenuator was adjusted until the capacitance obtained by the mSOL calibration differed by less than 5\,\% from the computed standard values, \cite{piquemal2021progress}. The attenuator is set to 7\,dB for $\it{f}_{VNA}$ below 6\,GHz and to 21\,dB for $\it{f}_{VNA}$ above 10\,GHz. LNE also used these images of MC2 samples to extract the gain of the matching network. \added{In particular, the capacitors with radii of approximately 2\,$\mu$m and 0.6\,$\mu$m and oxide heights of 50\,nm and 200\,nm are used for this. The University of Lille used an MC2 sample to determine the gain and did not account for the possible saturation of the VNA receivers. Two capacitors are used with radii of approximately 1.75\,$\mu$m and 0.7\,$\mu$m and oxide heights of 50\,nm and 400\,nm. The used capacitors are specified to simplify the reproduction of the results. Using capacitors that are too small in surface (capacitors below 100\,aF) may provoke false readings due to possible water meniscus and too large capacitors can be a problem for the dynamic range of the setup.}

In Table \ref{tabBestNoise} the best noise of each matching network at each institution is listed with $k=1$. This is the uncertainty contribution from the matching network which one would have to add on all measurement results. As said before, the translated mechanical noise and the noise from the VNA would have to be quadratically added to this for a more complete picture.

\begin{table}[h]
	\caption{Best Noise of Impedance Matching Setup}
  \label{tabBestNoise}
  \begin{center}
		\begin{tabular}{c | r }
      Setup & Noise in nS \\
      \hline
     METAS 5.395\,GHz & 500\\ 
     JKU 8\,GHz & 88 \\
     LNE 15.89\,GHz & 24\\
     Lille 4.465\,GHz & 5000
		\end{tabular}
  \end{center}
\end{table}

\section{Conclusion}
\added{The measured gain and noise values of the investigated impedance matching setups have uncertainties. In the case of noise, these uncertainties are mainly from limited repetitions of measurements and from the drift of the instrument during noise measurement. Typical noise uncertainties are 0.3\,dB with expansion factor $k=1$. The gain measurements suffer from uncertainties due to drift, noise, nonlinearity, and inaccurate calibration standards. In the case of the retraction curve technique, the determined gain suffers as well from drift, noise, and nonlinearity but may suffer additionally from synchronization inaccuracies between $z$-measurement and electric measurement. Typical gain uncertainties are 11\,\% with expansion factor $k=1$. This list of uncertainty influences is not exhaustive but should cover the most important ones. Better uncertainties of these values can be achieved by considering more points for the noise and gain measurements and thermally stabilized measurement setups.}

\added{This study is about two types of matching networks. The interferometric setups require two-port VNAs because a low noise amplifier is included in the measurement transmission chain. In addition, they require more sophisticated and expensive hardware such as continuously/stepped variable attenuators or/and phase-shifters. Consequently, they require more time to build and configure (as the level of amplitude of the microwave signal needs to be adjusted prior to microwave scanning). In contrast, the one-port setups (Beatty line and tuner) require less hardware, less time to build, and are less expensive.  On the other hand, the two-port networks offer more possibilities for setting signal amplitudes. This means one can set the step attenuators for a sample with very little or a lot of electric contrast such that the dynamic range of the VNA is fully used. This, in contrast, is not possible for the one-port techniques where the amplification is fixed. The effort for conducting measurements is similar for both types of setups, interferometric and one-port, but the effort for finding the right settings before measurement is higher for the interferometric setups. Another topic is the frequency flexibility. The interferometric setups cover a range of frequencies whereas the one-port setups cover only the frequencies mentioned in Tables \ref{tabGain} and \ref{tabNoise}. The achievable signal-to-noise ratios are approximately 25\,dB higher (JKU 127\,dB at 28\,GHz versus LNE 152\,dB at 15.89\,GHz) for the two-port techniques. Stability is another important factor for metrology applications. Typical stability values are better than -60\,dB per hour for interferometric setups and better than -75\,dB per hour for one-port setups. Here, it is better to have a setup without additional active elements and with as few components as possible. Thus, one can say that the more complicated interferometric setups with amplifiers have the benefit of adjustable gain and superior signal-to-noise ratios for short measurements. It has to be mentioned that the levels of cancellation signals and amplification need to be adjusted carefully as a function of the measurement environment. For example, a moderate amplification has to be considered in a non-controlled environment, as measured microwave signals are sensitive to temperature, humidity, and mechanical vibrations. The advantage of better signal-to-noise  vanishes if the measurement takes longer and comparisons have to be drawn between pixels that are temporally far apart. See as well Table \ref{tabQual} for a comparison between active and passive matching techniques.}

\begin{table}[h]
	\caption{Comparison Between Passive and Active  Impedance Matching Networks}
  \label{tabQual}
  \begin{center}
		\begin{tabular}{c| c | c }
       & passive & active \\
      \hline
      price & low &high \\
      VNA& one-port& two-port\\
      hardware requirements & low & high\\
      time to build setup & low & high \\
      time for measurement &same & same \\
      gain adjustable & no & yes \\
      frequency adjustable & single points & frequency range \\
      achieved SNR &127\,dB & 152\,dB \\
      stability & -75\,dB/h & -60\,dB/h \\
		\end{tabular}
  \end{center}
\end{table}

\section*{Acknowledgements}
This project (EMPIR 20IND12 ELENA) has received funding from the EMPIR programme co-financed by the Participating States and from the European Union's Horizon 2020 research and innovation programme. One of the authors thanks Alexis Litwin for technical support.

\bibliography{references}

\end{document}

%% file: Graphs/Setup_JKU.pdf_tex
\begingroup%
  \makeatletter%
  \providecommand\color[2][]{%
    \errmessage{(Inkscape) Color is used for the text in Inkscape, but the package 'color.sty' is not loaded}%
    \renewcommand\color[2][]{}%
  }%
  \providecommand\transparent[1]{%
    \errmessage{(Inkscape) Transparency is used (non-zero) for the text in Inkscape, but the package 'transparent.sty' is not loaded}%
    \renewcommand\transparent[1]{}%
  }%
  \providecommand\rotatebox[2]{#2}%
  \newcommand*\fsize{\dimexpr\f@size pt\relax}%
  \newcommand*\lineheight[1]{\fontsize{\fsize}{#1\fsize}\selectfont}%
  \ifx\svgwidth\undefined%
    \setlength{\unitlength}{520.69504505bp}%
    \ifx\svgscale\undefined%
      \relax%
    \else%
      \setlength{\unitlength}{\unitlength * \real{\svgscale}}%
    \fi%
  \else%
    \setlength{\unitlength}{\svgwidth}%
  \fi%
  \global\let\svgwidth\undefined%
  \global\let\svgscale\undefined%
  \makeatother%
  \begin{picture}(1,0.62520631)%
    \lineheight{1}%
    \setlength\tabcolsep{0pt}%
    \put(0,0){\includegraphics[width=\unitlength,page=1]{Setup_JKU.pdf}}%
    \put(0.25386247,0.41411103){\color[rgb]{0,0,0}\makebox(0,0)[t]{\lineheight{0.5}\smash{\begin{tabular}[t]{c}$S_{11}$\end{tabular}}}}%
    \put(0,0){\includegraphics[width=\unitlength,page=2]{Setup_JKU.pdf}}%
    \put(0.24624664,0.61061493){\color[rgb]{0,0,0}\makebox(0,0)[t]{\lineheight{1.25}\smash{\begin{tabular}[t]{c}Vector Network Analyzer\end{tabular}}}}%
    \put(0,0){\includegraphics[width=\unitlength,page=3]{Setup_JKU.pdf}}%
    \put(0.16990721,0.3670444){\color[rgb]{0,0,0}\makebox(0,0)[t]{\lineheight{0.5}\smash{\begin{tabular}[t]{c}\tiny Port 1\end{tabular}}}}%
    \put(0,0){\includegraphics[width=\unitlength,page=4]{Setup_JKU.pdf}}%
    \put(0.40036837,0.3670444){\color[rgb]{0,0,0}\makebox(0,0)[t]{\lineheight{0.5}\smash{\begin{tabular}[t]{c}\tiny Port 2\end{tabular}}}}%
    \put(0,0){\includegraphics[width=\unitlength,page=5]{Setup_JKU.pdf}}%
    \put(0.82327623,0.14462151){\color[rgb]{0,0,0}\makebox(0,0)[lt]{\lineheight{1.25}\smash{\begin{tabular}[t]{l}SMM Probe\end{tabular}}}}%
    \put(0,0){\includegraphics[width=\unitlength,page=6]{Setup_JKU.pdf}}%
    \put(0.43889105,0.22183251){\color[rgb]{0,0,0}\makebox(0,0)[lt]{\lineheight{1.25}\smash{\begin{tabular}[t]{l}Coaxial cable\end{tabular}}}}%
    \put(0,0){\includegraphics[width=\unitlength,page=7]{Setup_JKU.pdf}}%
    \put(0.34864809,0.12506963){\color[rgb]{0,0,0}\makebox(0,0)[t]{\lineheight{1.25}\smash{\begin{tabular}[t]{c}Impedance match\end{tabular}}}}%
    \put(0,0){\includegraphics[width=\unitlength,page=8]{Setup_JKU.pdf}}%
    \put(0.63518456,0.40432579){\color[rgb]{0,0,0}\makebox(0,0)[t]{\lineheight{1.25}\smash{\begin{tabular}[t]{c}Laser beam\end{tabular}}}}%
    \put(0.91582239,0.4069044){\color[rgb]{0,0,0}\makebox(0,0)[t]{\lineheight{1.25}\smash{\begin{tabular}[t]{c}Photodetector\end{tabular}}}}%
    \put(0,0){\includegraphics[width=\unitlength,page=9]{Setup_JKU.pdf}}%
    \put(0.69620147,0.10746293){\color[rgb]{0,0,0}\makebox(0,0)[t]{\lineheight{0.5}\smash{\begin{tabular}[t]{c}$a_{\mathrm{inc}}$\end{tabular}}}}%
    \put(0.85464141,0.10746293){\color[rgb]{0,0,0}\makebox(0,0)[t]{\lineheight{0.5}\smash{\begin{tabular}[t]{c}$b_{\mathrm{ref}}$\end{tabular}}}}%
    \put(0,0){\includegraphics[width=\unitlength,page=10]{Setup_JKU.pdf}}%
    \put(0.77517311,0.05421955){\color[rgb]{0,0,0}\makebox(0,0)[t]{\lineheight{1.25}\smash{\begin{tabular}[t]{c}Sample\end{tabular}}}}%
    \put(0,0){\includegraphics[width=\unitlength,page=11]{Setup_JKU.pdf}}%
  \end{picture}%
\endgroup%

%% file: Graphs/Setup_LNE.pdf_tex
\begingroup%
  \makeatletter%
  \providecommand\color[2][]{%
    \errmessage{(Inkscape) Color is used for the text in Inkscape, but the package 'color.sty' is not loaded}%
    \renewcommand\color[2][]{}%
  }%
  \providecommand\transparent[1]{%
    \errmessage{(Inkscape) Transparency is used (non-zero) for the text in Inkscape, but the package 'transparent.sty' is not loaded}%
    \renewcommand\transparent[1]{}%
  }%
  \providecommand\rotatebox[2]{#2}%
  \newcommand*\fsize{\dimexpr\f@size pt\relax}%
  \newcommand*\lineheight[1]{\fontsize{\fsize}{#1\fsize}\selectfont}%
  \ifx\svgwidth\undefined%
    \setlength{\unitlength}{544.3767263bp}%
    \ifx\svgscale\undefined%
      \relax%
    \else%
      \setlength{\unitlength}{\unitlength * \real{\svgscale}}%
    \fi%
  \else%
    \setlength{\unitlength}{\svgwidth}%
  \fi%
  \global\let\svgwidth\undefined%
  \global\let\svgscale\undefined%
  \makeatother%
  \begin{picture}(1,0.69239624)%
    \lineheight{1}%
    \setlength\tabcolsep{0pt}%
    \put(0,0){\includegraphics[width=\unitlength,page=1]{Setup_LNE.pdf}}%
    \put(0.78100913,0.48770986){\color[rgb]{0,0,0}\makebox(0,0)[t]{\lineheight{0.5}\smash{\begin{tabular}[t]{c}$S_{21}$ \end{tabular}}}}%
    \put(0.7590872,0.67843963){\color[rgb]{0,0,0}\makebox(0,0)[t]{\lineheight{1.25}\smash{\begin{tabular}[t]{c}Vector Network Analyzer\end{tabular}}}}%
    \put(0,0){\includegraphics[width=\unitlength,page=2]{Setup_LNE.pdf}}%
    \put(0.15009937,0.36782468){\color[rgb]{0,0,0}\makebox(0,0)[rt]{\lineheight{1.25}\smash{\begin{tabular}[t]{r}SMM Probe\end{tabular}}}}%
    \put(0.4396334,0.11754083){\color[rgb]{0,0,0}\makebox(0,0)[rt]{\lineheight{1.25}\smash{\begin{tabular}[t]{r}Coaxial cable\end{tabular}}}}%
    \put(0,0){\includegraphics[width=\unitlength,page=3]{Setup_LNE.pdf}}%
    \put(0.34873883,0.61623119){\color[rgb]{0,0,0}\makebox(0,0)[t]{\lineheight{1.25}\smash{\begin{tabular}[t]{c}Laser beam\end{tabular}}}}%
    \put(0.07866798,0.61869763){\color[rgb]{0,0,0}\makebox(0,0)[t]{\lineheight{1.25}\smash{\begin{tabular}[t]{c}Photodetector\end{tabular}}}}%
    \put(0,0){\includegraphics[width=\unitlength,page=4]{Setup_LNE.pdf}}%
    \put(0.29044445,0.33228255){\color[rgb]{0,0,0}\makebox(0,0)[t]{\lineheight{0.5}\smash{\begin{tabular}[t]{c}$a_{\mathrm{inc}}$\end{tabular}}}}%
    \put(0.13889702,0.33228255){\color[rgb]{0,0,0}\makebox(0,0)[t]{\lineheight{0.5}\smash{\begin{tabular}[t]{c}$b_{\mathrm{ref}}$\end{tabular}}}}%
    \put(0,0){\includegraphics[width=\unitlength,page=5]{Setup_LNE.pdf}}%
    \put(0.21474143,0.28135538){\color[rgb]{0,0,0}\makebox(0,0)[t]{\lineheight{1.25}\smash{\begin{tabular}[t]{c}Sample\end{tabular}}}}%
    \put(0,0){\includegraphics[width=\unitlength,page=6]{Setup_LNE.pdf}}%
    \put(0.98946499,0.01106191){\color[rgb]{0.37254902,0.55294118,0.82745098}\makebox(0,0)[rt]{\lineheight{1.25}\smash{\begin{tabular}[t]{r}Interferometer\end{tabular}}}}%
    \put(0,0){\includegraphics[width=\unitlength,page=7]{Setup_LNE.pdf}}%
    \put(0.62090774,0.26969739){\color[rgb]{0,0,0}\makebox(0,0)[rt]{\lineheight{1.25}\smash{\begin{tabular}[t]{r}Splitter\end{tabular}}}}%
    \put(0.87521596,0.26965254){\color[rgb]{0,0,0}\makebox(0,0)[rt]{\lineheight{1.25}\smash{\begin{tabular}[t]{r}Amplifier\end{tabular}}}}%
    \put(0.7564818,0.15906097){\color[rgb]{0,0,0}\makebox(0,0)[lt]{\lineheight{1.25}\smash{\begin{tabular}[t]{l}Attenuator\end{tabular}}}}%
    \put(0.59137203,0.04606657){\color[rgb]{0,0,0}\makebox(0,0)[lt]{\lineheight{1.25}\smash{\begin{tabular}[t]{l}Directional couplers\end{tabular}}}}%
    \put(0,0){\includegraphics[width=\unitlength,page=8]{Setup_LNE.pdf}}%
    \put(0.8983066,0.48783542){\color[rgb]{0,0,0}\makebox(0,0)[lt]{\lineheight{1.25}\smash{\begin{tabular}[t]{l}$\color{red}(S_{11})$\end{tabular}}}}%
    \put(0.59324136,0.45476602){\color[rgb]{0,0,0}\makebox(0,0)[t]{\lineheight{0.5}\smash{\begin{tabular}[t]{c}\tiny Rec. A\end{tabular}}}}%
    \put(0.59290721,0.42174089){\color[rgb]{0,0,0}\makebox(0,0)[t]{\lineheight{0.5}\smash{\begin{tabular}[t]{c}\tiny Source\end{tabular}}}}%
    \put(0,0){\includegraphics[width=\unitlength,page=9]{Setup_LNE.pdf}}%
    \put(0.90394128,0.44283761){\color[rgb]{0,0,0}\makebox(0,0)[t]{\lineheight{0.5}\smash{\begin{tabular}[t]{c}\tiny Port 2\end{tabular}}}}%
    \put(0,0){\includegraphics[width=\unitlength,page=10]{Setup_LNE.pdf}}%
    \put(0.68350517,0.44283761){\color[rgb]{0,0,0}\makebox(0,0)[t]{\lineheight{0.5}\smash{\begin{tabular}[t]{c}\tiny Port 1\end{tabular}}}}%
    \put(0,0){\includegraphics[width=\unitlength,page=11]{Setup_LNE.pdf}}%
  \end{picture}%
\endgroup%

%% file: Graphs/Setup_Lille.pdf_tex
\begingroup%
  \makeatletter%
  \providecommand\color[2][]{%
    \errmessage{(Inkscape) Color is used for the text in Inkscape, but the package 'color.sty' is not loaded}%
    \renewcommand\color[2][]{}%
  }%
  \providecommand\transparent[1]{%
    \errmessage{(Inkscape) Transparency is used (non-zero) for the text in Inkscape, but the package 'transparent.sty' is not loaded}%
    \renewcommand\transparent[1]{}%
  }%
  \providecommand\rotatebox[2]{#2}%
  \newcommand*\fsize{\dimexpr\f@size pt\relax}%
  \newcommand*\lineheight[1]{\fontsize{\fsize}{#1\fsize}\selectfont}%
  \ifx\svgwidth\undefined%
    \setlength{\unitlength}{278.8860195bp}%
    \ifx\svgscale\undefined%
      \relax%
    \else%
      \setlength{\unitlength}{\unitlength * \real{\svgscale}}%
    \fi%
  \else%
    \setlength{\unitlength}{\svgwidth}%
  \fi%
  \global\let\svgwidth\undefined%
  \global\let\svgscale\undefined%
  \makeatother%
  \begin{picture}(1,1.5073256)%
    \lineheight{1}%
    \setlength\tabcolsep{0pt}%
    \put(0,0){\includegraphics[width=\unitlength,page=1]{Setup_Lille.pdf}}%
    \put(0.56033982,0.75736001){\color[rgb]{0,0,0}\makebox(0,0)[t]{\lineheight{1.25}\smash{\begin{tabular}[t]{c}Source\end{tabular}}}}%
    \put(0.31946277,0.63940708){\color[rgb]{0,0,0}\makebox(0,0)[lt]{\lineheight{0.1}\smash{\begin{tabular}[t]{l}Reference\\receiver\end{tabular}}}}%
    \put(0.74458833,0.59429643){\color[rgb]{0,0,0}\makebox(0,0)[t]{\lineheight{1.25}\smash{\begin{tabular}[t]{c}Measurement receiver\end{tabular}}}}%
    \put(0,0){\includegraphics[width=\unitlength,page=2]{Setup_Lille.pdf}}%
    \put(0.27394399,0.61553602){\color[rgb]{0,0,0}\makebox(0,0)[t]{\lineheight{1.25}\smash{\begin{tabular}[t]{c}R1\end{tabular}}}}%
    \put(0,0){\includegraphics[width=\unitlength,page=3]{Setup_Lille.pdf}}%
    \put(0.70518782,0.51632092){\color[rgb]{0,0,0}\makebox(0,0)[t]{\lineheight{1.25}\smash{\begin{tabular}[t]{c}A\end{tabular}}}}%
    \put(0,0){\includegraphics[width=\unitlength,page=4]{Setup_Lille.pdf}}%
    \put(0.70518782,0.51632092){\color[rgb]{0,0,0}\makebox(0,0)[t]{\lineheight{1.25}\smash{\begin{tabular}[t]{c}A\end{tabular}}}}%
    \put(0,0){\includegraphics[width=\unitlength,page=5]{Setup_Lille.pdf}}%
    \put(0.53963088,1.14005551){\color[rgb]{0,0,0}\makebox(0,0)[t]{\lineheight{0.5}\smash{\begin{tabular}[t]{c}$S_{11}$ \end{tabular}}}}%
    \put(0.49683996,1.48008273){\color[rgb]{0,0,0}\makebox(0,0)[t]{\lineheight{1.25}\smash{\begin{tabular}[t]{c}Vector Network Analyzer\end{tabular}}}}%
    \put(0.16235704,1.0757502){\color[rgb]{0,0,0}\makebox(0,0)[t]{\lineheight{0.5}\smash{\begin{tabular}[t]{c}\tiny Rec. A\end{tabular}}}}%
    \put(0,0){\includegraphics[width=\unitlength,page=6]{Setup_Lille.pdf}}%
    \put(0.77959064,1.05246628){\color[rgb]{0,0,0}\makebox(0,0)[t]{\lineheight{0.5}\smash{\begin{tabular}[t]{c}\tiny Port 2\end{tabular}}}}%
    \put(0,0){\includegraphics[width=\unitlength,page=7]{Setup_Lille.pdf}}%
    \put(0.34930622,1.05246628){\color[rgb]{0,0,0}\makebox(0,0)[t]{\lineheight{0.5}\smash{\begin{tabular}[t]{c}\tiny Port 1\end{tabular}}}}%
    \put(0.1617048,0.94136511){\color[rgb]{0,0,0}\makebox(0,0)[t]{\lineheight{0.5}\smash{\begin{tabular}[t]{c}\tiny Ref.\end{tabular}}}}%
    \put(0,0){\includegraphics[width=\unitlength,page=8]{Setup_Lille.pdf}}%
    \put(0.74542343,0.03771449){\color[rgb]{0.37254902,0.55294118,0.82745098}\makebox(0,0)[rt]{\lineheight{1.25}\smash{\begin{tabular}[t]{r}Interferometer\end{tabular}}}}%
    \put(0,0){\includegraphics[width=\unitlength,page=9]{Setup_Lille.pdf}}%
    \put(0.22104263,0.11531115){\color[rgb]{0,0,0}\makebox(0,0)[rt]{\lineheight{1.25}\smash{\begin{tabular}[t]{r}Splitter\end{tabular}}}}%
    \put(0.71744349,0.1152236){\color[rgb]{0,0,0}\makebox(0,0)[rt]{\lineheight{1.25}\smash{\begin{tabular}[t]{r}Amplifier\end{tabular}}}}%
    \put(0,0){\includegraphics[width=\unitlength,page=10]{Setup_Lille.pdf}}%
    \put(0.1617048,1.01128614){\color[rgb]{0,0,0}\makebox(0,0)[t]{\lineheight{0.5}\smash{\begin{tabular}[t]{c}\tiny Source\end{tabular}}}}%
    \put(0,0){\includegraphics[width=\unitlength,page=11]{Setup_Lille.pdf}}%
  \end{picture}%
\endgroup%

%% file: abs_s11.pgf
\begingroup%
\makeatletter%
\begin{pgfpicture}%
\pgfpathrectangle{\pgfpointorigin}{\pgfqpoint{5.513637in}{4.355710in}}%
\pgfusepath{use as bounding box, clip}%
\begin{pgfscope}%
\pgfsetbuttcap%
\pgfsetmiterjoin%
\definecolor{currentfill}{rgb}{1.000000,1.000000,1.000000}%
\pgfsetfillcolor{currentfill}%
\pgfsetlinewidth{0.000000pt}%
\definecolor{currentstroke}{rgb}{1.000000,1.000000,1.000000}%
\pgfsetstrokecolor{currentstroke}%
\pgfsetdash{}{0pt}%
\pgfpathmoveto{\pgfqpoint{0.000000in}{0.000000in}}%
\pgfpathlineto{\pgfqpoint{5.513637in}{0.000000in}}%
\pgfpathlineto{\pgfqpoint{5.513637in}{4.355710in}}%
\pgfpathlineto{\pgfqpoint{0.000000in}{4.355710in}}%
\pgfpathclose%
\pgfusepath{fill}%
\end{pgfscope}%
\begin{pgfscope}%
\pgfsetbuttcap%
\pgfsetmiterjoin%
\definecolor{currentfill}{rgb}{1.000000,1.000000,1.000000}%
\pgfsetfillcolor{currentfill}%
\pgfsetlinewidth{0.000000pt}%
\definecolor{currentstroke}{rgb}{0.000000,0.000000,0.000000}%
\pgfsetstrokecolor{currentstroke}%
\pgfsetstrokeopacity{0.000000}%
\pgfsetdash{}{0pt}%
\pgfpathmoveto{\pgfqpoint{0.360415in}{0.345370in}}%
\pgfpathlineto{\pgfqpoint{4.760415in}{0.345370in}}%
\pgfpathlineto{\pgfqpoint{4.760415in}{4.041370in}}%
\pgfpathlineto{\pgfqpoint{0.360415in}{4.041370in}}%
\pgfpathclose%
\pgfusepath{fill}%
\end{pgfscope}%
\begin{pgfscope}%
\pgfpathrectangle{\pgfqpoint{0.360415in}{0.345370in}}{\pgfqpoint{4.400000in}{3.696000in}}%
\pgfusepath{clip}%
\pgfsys@transformcm{4.400000}{0.000000}{0.000000}{-3.700000}{0.360415in}{4.045370in}%
\pgftext[left,bottom]{\pgfimage[interpolate=false,width=1.000000in,height=1.000000in]{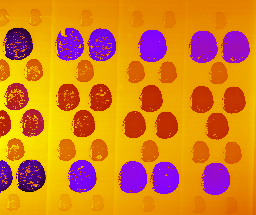}}%
\end{pgfscope}%
\begin{pgfscope}%
\pgfsetbuttcap%
\pgfsetroundjoin%
\definecolor{currentfill}{rgb}{0.000000,0.000000,0.000000}%
\pgfsetfillcolor{currentfill}%
\pgfsetlinewidth{0.803000pt}%
\definecolor{currentstroke}{rgb}{0.000000,0.000000,0.000000}%
\pgfsetstrokecolor{currentstroke}%
\pgfsetdash{}{0pt}%
\pgfsys@defobject{currentmarker}{\pgfqpoint{0.000000in}{-0.048611in}}{\pgfqpoint{0.000000in}{0.000000in}}{%
\pgfpathmoveto{\pgfqpoint{0.000000in}{0.000000in}}%
\pgfpathlineto{\pgfqpoint{0.000000in}{-0.048611in}}%
\pgfusepath{stroke,fill}%
}%
\begin{pgfscope}%
\pgfsys@transformshift{0.360415in}{0.345370in}%
\pgfsys@useobject{currentmarker}{}%
\end{pgfscope}%
\end{pgfscope}%
\begin{pgfscope}%
\definecolor{textcolor}{rgb}{0.000000,0.000000,0.000000}%
\pgfsetstrokecolor{textcolor}%
\pgfsetfillcolor{textcolor}%
\pgftext[x=0.360415in,y=0.248148in,,top]{\color{textcolor}\sffamily\fontsize{12.000000}{14.400000}\selectfont \(\displaystyle 0\)}%
\end{pgfscope}%
\begin{pgfscope}%
\pgfsetbuttcap%
\pgfsetroundjoin%
\definecolor{currentfill}{rgb}{0.000000,0.000000,0.000000}%
\pgfsetfillcolor{currentfill}%
\pgfsetlinewidth{0.803000pt}%
\definecolor{currentstroke}{rgb}{0.000000,0.000000,0.000000}%
\pgfsetstrokecolor{currentstroke}%
\pgfsetdash{}{0pt}%
\pgfsys@defobject{currentmarker}{\pgfqpoint{0.000000in}{-0.048611in}}{\pgfqpoint{0.000000in}{0.000000in}}{%
\pgfpathmoveto{\pgfqpoint{0.000000in}{0.000000in}}%
\pgfpathlineto{\pgfqpoint{0.000000in}{-0.048611in}}%
\pgfusepath{stroke,fill}%
}%
\begin{pgfscope}%
\pgfsys@transformshift{1.240415in}{0.345370in}%
\pgfsys@useobject{currentmarker}{}%
\end{pgfscope}%
\end{pgfscope}%
\begin{pgfscope}%
\definecolor{textcolor}{rgb}{0.000000,0.000000,0.000000}%
\pgfsetstrokecolor{textcolor}%
\pgfsetfillcolor{textcolor}%
\pgftext[x=1.240415in,y=0.248148in,,top]{\color{textcolor}\sffamily\fontsize{12.000000}{14.400000}\selectfont \(\displaystyle 10\)}%
\end{pgfscope}%
\begin{pgfscope}%
\pgfsetbuttcap%
\pgfsetroundjoin%
\definecolor{currentfill}{rgb}{0.000000,0.000000,0.000000}%
\pgfsetfillcolor{currentfill}%
\pgfsetlinewidth{0.803000pt}%
\definecolor{currentstroke}{rgb}{0.000000,0.000000,0.000000}%
\pgfsetstrokecolor{currentstroke}%
\pgfsetdash{}{0pt}%
\pgfsys@defobject{currentmarker}{\pgfqpoint{0.000000in}{-0.048611in}}{\pgfqpoint{0.000000in}{0.000000in}}{%
\pgfpathmoveto{\pgfqpoint{0.000000in}{0.000000in}}%
\pgfpathlineto{\pgfqpoint{0.000000in}{-0.048611in}}%
\pgfusepath{stroke,fill}%
}%
\begin{pgfscope}%
\pgfsys@transformshift{2.120415in}{0.345370in}%
\pgfsys@useobject{currentmarker}{}%
\end{pgfscope}%
\end{pgfscope}%
\begin{pgfscope}%
\definecolor{textcolor}{rgb}{0.000000,0.000000,0.000000}%
\pgfsetstrokecolor{textcolor}%
\pgfsetfillcolor{textcolor}%
\pgftext[x=2.120415in,y=0.248148in,,top]{\color{textcolor}\sffamily\fontsize{12.000000}{14.400000}\selectfont \(\displaystyle 20\)}%
\end{pgfscope}%
\begin{pgfscope}%
\pgfsetbuttcap%
\pgfsetroundjoin%
\definecolor{currentfill}{rgb}{0.000000,0.000000,0.000000}%
\pgfsetfillcolor{currentfill}%
\pgfsetlinewidth{0.803000pt}%
\definecolor{currentstroke}{rgb}{0.000000,0.000000,0.000000}%
\pgfsetstrokecolor{currentstroke}%
\pgfsetdash{}{0pt}%
\pgfsys@defobject{currentmarker}{\pgfqpoint{0.000000in}{-0.048611in}}{\pgfqpoint{0.000000in}{0.000000in}}{%
\pgfpathmoveto{\pgfqpoint{0.000000in}{0.000000in}}%
\pgfpathlineto{\pgfqpoint{0.000000in}{-0.048611in}}%
\pgfusepath{stroke,fill}%
}%
\begin{pgfscope}%
\pgfsys@transformshift{3.000415in}{0.345370in}%
\pgfsys@useobject{currentmarker}{}%
\end{pgfscope}%
\end{pgfscope}%
\begin{pgfscope}%
\definecolor{textcolor}{rgb}{0.000000,0.000000,0.000000}%
\pgfsetstrokecolor{textcolor}%
\pgfsetfillcolor{textcolor}%
\pgftext[x=3.000415in,y=0.248148in,,top]{\color{textcolor}\sffamily\fontsize{12.000000}{14.400000}\selectfont \(\displaystyle 30\)}%
\end{pgfscope}%
\begin{pgfscope}%
\pgfsetbuttcap%
\pgfsetroundjoin%
\definecolor{currentfill}{rgb}{0.000000,0.000000,0.000000}%
\pgfsetfillcolor{currentfill}%
\pgfsetlinewidth{0.803000pt}%
\definecolor{currentstroke}{rgb}{0.000000,0.000000,0.000000}%
\pgfsetstrokecolor{currentstroke}%
\pgfsetdash{}{0pt}%
\pgfsys@defobject{currentmarker}{\pgfqpoint{0.000000in}{-0.048611in}}{\pgfqpoint{0.000000in}{0.000000in}}{%
\pgfpathmoveto{\pgfqpoint{0.000000in}{0.000000in}}%
\pgfpathlineto{\pgfqpoint{0.000000in}{-0.048611in}}%
\pgfusepath{stroke,fill}%
}%
\begin{pgfscope}%
\pgfsys@transformshift{3.880415in}{0.345370in}%
\pgfsys@useobject{currentmarker}{}%
\end{pgfscope}%
\end{pgfscope}%
\begin{pgfscope}%
\definecolor{textcolor}{rgb}{0.000000,0.000000,0.000000}%
\pgfsetstrokecolor{textcolor}%
\pgfsetfillcolor{textcolor}%
\pgftext[x=3.880415in,y=0.248148in,,top]{\color{textcolor}\sffamily\fontsize{12.000000}{14.400000}\selectfont \(\displaystyle 40\)}%
\end{pgfscope}%
\begin{pgfscope}%
\pgfsetbuttcap%
\pgfsetroundjoin%
\definecolor{currentfill}{rgb}{0.000000,0.000000,0.000000}%
\pgfsetfillcolor{currentfill}%
\pgfsetlinewidth{0.803000pt}%
\definecolor{currentstroke}{rgb}{0.000000,0.000000,0.000000}%
\pgfsetstrokecolor{currentstroke}%
\pgfsetdash{}{0pt}%
\pgfsys@defobject{currentmarker}{\pgfqpoint{0.000000in}{-0.048611in}}{\pgfqpoint{0.000000in}{0.000000in}}{%
\pgfpathmoveto{\pgfqpoint{0.000000in}{0.000000in}}%
\pgfpathlineto{\pgfqpoint{0.000000in}{-0.048611in}}%
\pgfusepath{stroke,fill}%
}%
\begin{pgfscope}%
\pgfsys@transformshift{4.760415in}{0.345370in}%
\pgfsys@useobject{currentmarker}{}%
\end{pgfscope}%
\end{pgfscope}%
\begin{pgfscope}%
\definecolor{textcolor}{rgb}{0.000000,0.000000,0.000000}%
\pgfsetstrokecolor{textcolor}%
\pgfsetfillcolor{textcolor}%
\pgftext[x=4.760415in,y=0.248148in,,top]{\color{textcolor}\sffamily\fontsize{12.000000}{14.400000}\selectfont \(\displaystyle 50\)}%
\end{pgfscope}%
\begin{pgfscope}%
\pgfsetbuttcap%
\pgfsetroundjoin%
\definecolor{currentfill}{rgb}{0.000000,0.000000,0.000000}%
\pgfsetfillcolor{currentfill}%
\pgfsetlinewidth{0.803000pt}%
\definecolor{currentstroke}{rgb}{0.000000,0.000000,0.000000}%
\pgfsetstrokecolor{currentstroke}%
\pgfsetdash{}{0pt}%
\pgfsys@defobject{currentmarker}{\pgfqpoint{-0.048611in}{0.000000in}}{\pgfqpoint{0.000000in}{0.000000in}}{%
\pgfpathmoveto{\pgfqpoint{0.000000in}{0.000000in}}%
\pgfpathlineto{\pgfqpoint{-0.048611in}{0.000000in}}%
\pgfusepath{stroke,fill}%
}%
\begin{pgfscope}%
\pgfsys@transformshift{0.360415in}{0.345370in}%
\pgfsys@useobject{currentmarker}{}%
\end{pgfscope}%
\end{pgfscope}%
\begin{pgfscope}%
\definecolor{textcolor}{rgb}{0.000000,0.000000,0.000000}%
\pgfsetstrokecolor{textcolor}%
\pgfsetfillcolor{textcolor}%
\pgftext[x=0.181596in,y=0.287500in,left,base]{\color{textcolor}\sffamily\fontsize{12.000000}{14.400000}\selectfont \(\displaystyle 0\)}%
\end{pgfscope}%
\begin{pgfscope}%
\pgfsetbuttcap%
\pgfsetroundjoin%
\definecolor{currentfill}{rgb}{0.000000,0.000000,0.000000}%
\pgfsetfillcolor{currentfill}%
\pgfsetlinewidth{0.803000pt}%
\definecolor{currentstroke}{rgb}{0.000000,0.000000,0.000000}%
\pgfsetstrokecolor{currentstroke}%
\pgfsetdash{}{0pt}%
\pgfsys@defobject{currentmarker}{\pgfqpoint{-0.048611in}{0.000000in}}{\pgfqpoint{0.000000in}{0.000000in}}{%
\pgfpathmoveto{\pgfqpoint{0.000000in}{0.000000in}}%
\pgfpathlineto{\pgfqpoint{-0.048611in}{0.000000in}}%
\pgfusepath{stroke,fill}%
}%
\begin{pgfscope}%
\pgfsys@transformshift{0.360415in}{1.225370in}%
\pgfsys@useobject{currentmarker}{}%
\end{pgfscope}%
\end{pgfscope}%
\begin{pgfscope}%
\definecolor{textcolor}{rgb}{0.000000,0.000000,0.000000}%
\pgfsetstrokecolor{textcolor}%
\pgfsetfillcolor{textcolor}%
\pgftext[x=0.100000in,y=1.167500in,left,base]{\color{textcolor}\sffamily\fontsize{12.000000}{14.400000}\selectfont \(\displaystyle 10\)}%
\end{pgfscope}%
\begin{pgfscope}%
\pgfsetbuttcap%
\pgfsetroundjoin%
\definecolor{currentfill}{rgb}{0.000000,0.000000,0.000000}%
\pgfsetfillcolor{currentfill}%
\pgfsetlinewidth{0.803000pt}%
\definecolor{currentstroke}{rgb}{0.000000,0.000000,0.000000}%
\pgfsetstrokecolor{currentstroke}%
\pgfsetdash{}{0pt}%
\pgfsys@defobject{currentmarker}{\pgfqpoint{-0.048611in}{0.000000in}}{\pgfqpoint{0.000000in}{0.000000in}}{%
\pgfpathmoveto{\pgfqpoint{0.000000in}{0.000000in}}%
\pgfpathlineto{\pgfqpoint{-0.048611in}{0.000000in}}%
\pgfusepath{stroke,fill}%
}%
\begin{pgfscope}%
\pgfsys@transformshift{0.360415in}{2.105370in}%
\pgfsys@useobject{currentmarker}{}%
\end{pgfscope}%
\end{pgfscope}%
\begin{pgfscope}%
\definecolor{textcolor}{rgb}{0.000000,0.000000,0.000000}%
\pgfsetstrokecolor{textcolor}%
\pgfsetfillcolor{textcolor}%
\pgftext[x=0.100000in,y=2.047500in,left,base]{\color{textcolor}\sffamily\fontsize{12.000000}{14.400000}\selectfont \(\displaystyle 20\)}%
\end{pgfscope}%
\begin{pgfscope}%
\pgfsetbuttcap%
\pgfsetroundjoin%
\definecolor{currentfill}{rgb}{0.000000,0.000000,0.000000}%
\pgfsetfillcolor{currentfill}%
\pgfsetlinewidth{0.803000pt}%
\definecolor{currentstroke}{rgb}{0.000000,0.000000,0.000000}%
\pgfsetstrokecolor{currentstroke}%
\pgfsetdash{}{0pt}%
\pgfsys@defobject{currentmarker}{\pgfqpoint{-0.048611in}{0.000000in}}{\pgfqpoint{0.000000in}{0.000000in}}{%
\pgfpathmoveto{\pgfqpoint{0.000000in}{0.000000in}}%
\pgfpathlineto{\pgfqpoint{-0.048611in}{0.000000in}}%
\pgfusepath{stroke,fill}%
}%
\begin{pgfscope}%
\pgfsys@transformshift{0.360415in}{2.985370in}%
\pgfsys@useobject{currentmarker}{}%
\end{pgfscope}%
\end{pgfscope}%
\begin{pgfscope}%
\definecolor{textcolor}{rgb}{0.000000,0.000000,0.000000}%
\pgfsetstrokecolor{textcolor}%
\pgfsetfillcolor{textcolor}%
\pgftext[x=0.100000in,y=2.927500in,left,base]{\color{textcolor}\sffamily\fontsize{12.000000}{14.400000}\selectfont \(\displaystyle 30\)}%
\end{pgfscope}%
\begin{pgfscope}%
\pgfsetbuttcap%
\pgfsetroundjoin%
\definecolor{currentfill}{rgb}{0.000000,0.000000,0.000000}%
\pgfsetfillcolor{currentfill}%
\pgfsetlinewidth{0.803000pt}%
\definecolor{currentstroke}{rgb}{0.000000,0.000000,0.000000}%
\pgfsetstrokecolor{currentstroke}%
\pgfsetdash{}{0pt}%
\pgfsys@defobject{currentmarker}{\pgfqpoint{-0.048611in}{0.000000in}}{\pgfqpoint{0.000000in}{0.000000in}}{%
\pgfpathmoveto{\pgfqpoint{0.000000in}{0.000000in}}%
\pgfpathlineto{\pgfqpoint{-0.048611in}{0.000000in}}%
\pgfusepath{stroke,fill}%
}%
\begin{pgfscope}%
\pgfsys@transformshift{0.360415in}{3.865370in}%
\pgfsys@useobject{currentmarker}{}%
\end{pgfscope}%
\end{pgfscope}%
\begin{pgfscope}%
\definecolor{textcolor}{rgb}{0.000000,0.000000,0.000000}%
\pgfsetstrokecolor{textcolor}%
\pgfsetfillcolor{textcolor}%
\pgftext[x=0.100000in,y=3.807500in,left,base]{\color{textcolor}\sffamily\fontsize{12.000000}{14.400000}\selectfont \(\displaystyle 40\)}%
\end{pgfscope}%
\begin{pgfscope}%
\pgfsetrectcap%
\pgfsetmiterjoin%
\pgfsetlinewidth{0.803000pt}%
\definecolor{currentstroke}{rgb}{0.000000,0.000000,0.000000}%
\pgfsetstrokecolor{currentstroke}%
\pgfsetdash{}{0pt}%
\pgfpathmoveto{\pgfqpoint{0.360415in}{0.345370in}}%
\pgfpathlineto{\pgfqpoint{0.360415in}{4.041370in}}%
\pgfusepath{stroke}%
\end{pgfscope}%
\begin{pgfscope}%
\pgfsetrectcap%
\pgfsetmiterjoin%
\pgfsetlinewidth{0.803000pt}%
\definecolor{currentstroke}{rgb}{0.000000,0.000000,0.000000}%
\pgfsetstrokecolor{currentstroke}%
\pgfsetdash{}{0pt}%
\pgfpathmoveto{\pgfqpoint{4.760415in}{0.345370in}}%
\pgfpathlineto{\pgfqpoint{4.760415in}{4.041370in}}%
\pgfusepath{stroke}%
\end{pgfscope}%
\begin{pgfscope}%
\pgfsetrectcap%
\pgfsetmiterjoin%
\pgfsetlinewidth{0.803000pt}%
\definecolor{currentstroke}{rgb}{0.000000,0.000000,0.000000}%
\pgfsetstrokecolor{currentstroke}%
\pgfsetdash{}{0pt}%
\pgfpathmoveto{\pgfqpoint{0.360415in}{0.345370in}}%
\pgfpathlineto{\pgfqpoint{4.760415in}{0.345370in}}%
\pgfusepath{stroke}%
\end{pgfscope}%
\begin{pgfscope}%
\pgfsetrectcap%
\pgfsetmiterjoin%
\pgfsetlinewidth{0.803000pt}%
\definecolor{currentstroke}{rgb}{0.000000,0.000000,0.000000}%
\pgfsetstrokecolor{currentstroke}%
\pgfsetdash{}{0pt}%
\pgfpathmoveto{\pgfqpoint{0.360415in}{4.041370in}}%
\pgfpathlineto{\pgfqpoint{4.760415in}{4.041370in}}%
\pgfusepath{stroke}%
\end{pgfscope}%
\begin{pgfscope}%
\definecolor{textcolor}{rgb}{0.000000,0.000000,0.000000}%
\pgfsetstrokecolor{textcolor}%
\pgfsetfillcolor{textcolor}%
\pgftext[x=4.540415in,y=4.115290in,left,base]{\color{textcolor}\sffamily\fontsize{12.000000}{14.400000}\selectfont \(\displaystyle |S11| \cdot 10^3\)}%
\end{pgfscope}%
\begin{pgfscope}%
\definecolor{textcolor}{rgb}{1.000000,1.000000,1.000000}%
\pgfsetstrokecolor{textcolor}%
\pgfsetfillcolor{textcolor}%
\pgftext[x=0.897215in,y=3.372570in,,]{\color{textcolor}\sffamily\fontsize{18.000000}{21.600000}\selectfont \(\displaystyle \mathbf{C_1}\)}%
\end{pgfscope}%
\begin{pgfscope}%
\definecolor{textcolor}{rgb}{1.000000,1.000000,1.000000}%
\pgfsetstrokecolor{textcolor}%
\pgfsetfillcolor{textcolor}%
\pgftext[x=4.338015in,y=2.985370in,,]{\color{textcolor}\sffamily\fontsize{18.000000}{21.600000}\selectfont \(\displaystyle \mathbf{C_2}\)}%
\end{pgfscope}%
\begin{pgfscope}%
\definecolor{textcolor}{rgb}{0.000000,0.000000,0.000000}%
\pgfsetstrokecolor{textcolor}%
\pgfsetfillcolor{textcolor}%
\pgftext[x=0.448415in,y=0.142090in,left,base]{\color{textcolor}\sffamily\fontsize{12.000000}{14.400000}\selectfont \(\displaystyle \mu m\)}%
\end{pgfscope}%
\begin{pgfscope}%
\pgfpathrectangle{\pgfqpoint{4.810415in}{0.345370in}}{\pgfqpoint{0.220000in}{3.696000in}}%
\pgfusepath{clip}%
\pgfsetbuttcap%
\pgfsetmiterjoin%
\definecolor{currentfill}{rgb}{1.000000,1.000000,1.000000}%
\pgfsetfillcolor{currentfill}%
\pgfsetlinewidth{0.010037pt}%
\definecolor{currentstroke}{rgb}{1.000000,1.000000,1.000000}%
\pgfsetstrokecolor{currentstroke}%
\pgfsetdash{}{0pt}%
\pgfpathmoveto{\pgfqpoint{4.810415in}{0.345370in}}%
\pgfpathlineto{\pgfqpoint{4.810415in}{0.359807in}}%
\pgfpathlineto{\pgfqpoint{4.810415in}{4.026932in}}%
\pgfpathlineto{\pgfqpoint{4.810415in}{4.041370in}}%
\pgfpathlineto{\pgfqpoint{5.030415in}{4.041370in}}%
\pgfpathlineto{\pgfqpoint{5.030415in}{4.026932in}}%
\pgfpathlineto{\pgfqpoint{5.030415in}{0.359807in}}%
\pgfpathlineto{\pgfqpoint{5.030415in}{0.345370in}}%
\pgfpathclose%
\pgfusepath{stroke,fill}%
\end{pgfscope}%
\begin{pgfscope}%
\pgfsys@transformshift{4.810000in}{0.355710in}%
\pgftext[left,bottom]{\pgfimage[interpolate=true,width=0.220000in,height=3.690000in]{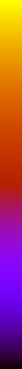}}%
\end{pgfscope}%
\begin{pgfscope}%
\pgfsetbuttcap%
\pgfsetroundjoin%
\definecolor{currentfill}{rgb}{0.000000,0.000000,0.000000}%
\pgfsetfillcolor{currentfill}%
\pgfsetlinewidth{0.803000pt}%
\definecolor{currentstroke}{rgb}{0.000000,0.000000,0.000000}%
\pgfsetstrokecolor{currentstroke}%
\pgfsetdash{}{0pt}%
\pgfsys@defobject{currentmarker}{\pgfqpoint{0.000000in}{0.000000in}}{\pgfqpoint{0.048611in}{0.000000in}}{%
\pgfpathmoveto{\pgfqpoint{0.000000in}{0.000000in}}%
\pgfpathlineto{\pgfqpoint{0.048611in}{0.000000in}}%
\pgfusepath{stroke,fill}%
}%
\begin{pgfscope}%
\pgfsys@transformshift{5.030415in}{0.671462in}%
\pgfsys@useobject{currentmarker}{}%
\end{pgfscope}%
\end{pgfscope}%
\begin{pgfscope}%
\definecolor{textcolor}{rgb}{0.000000,0.000000,0.000000}%
\pgfsetstrokecolor{textcolor}%
\pgfsetfillcolor{textcolor}%
\pgftext[x=5.075483in,y=0.613592in,left,base]{\color{textcolor}\sffamily\fontsize{12.000000}{14.400000}\selectfont \(\displaystyle -2.5\)}%
\end{pgfscope}%
\begin{pgfscope}%
\pgfsetbuttcap%
\pgfsetroundjoin%
\definecolor{currentfill}{rgb}{0.000000,0.000000,0.000000}%
\pgfsetfillcolor{currentfill}%
\pgfsetlinewidth{0.803000pt}%
\definecolor{currentstroke}{rgb}{0.000000,0.000000,0.000000}%
\pgfsetstrokecolor{currentstroke}%
\pgfsetdash{}{0pt}%
\pgfsys@defobject{currentmarker}{\pgfqpoint{0.000000in}{0.000000in}}{\pgfqpoint{0.048611in}{0.000000in}}{%
\pgfpathmoveto{\pgfqpoint{0.000000in}{0.000000in}}%
\pgfpathlineto{\pgfqpoint{0.048611in}{0.000000in}}%
\pgfusepath{stroke,fill}%
}%
\begin{pgfscope}%
\pgfsys@transformshift{5.030415in}{1.152745in}%
\pgfsys@useobject{currentmarker}{}%
\end{pgfscope}%
\end{pgfscope}%
\begin{pgfscope}%
\definecolor{textcolor}{rgb}{0.000000,0.000000,0.000000}%
\pgfsetstrokecolor{textcolor}%
\pgfsetfillcolor{textcolor}%
\pgftext[x=5.075483in,y=1.094875in,left,base]{\color{textcolor}\sffamily\fontsize{12.000000}{14.400000}\selectfont \(\displaystyle -2.0\)}%
\end{pgfscope}%
\begin{pgfscope}%
\pgfsetbuttcap%
\pgfsetroundjoin%
\definecolor{currentfill}{rgb}{0.000000,0.000000,0.000000}%
\pgfsetfillcolor{currentfill}%
\pgfsetlinewidth{0.803000pt}%
\definecolor{currentstroke}{rgb}{0.000000,0.000000,0.000000}%
\pgfsetstrokecolor{currentstroke}%
\pgfsetdash{}{0pt}%
\pgfsys@defobject{currentmarker}{\pgfqpoint{0.000000in}{0.000000in}}{\pgfqpoint{0.048611in}{0.000000in}}{%
\pgfpathmoveto{\pgfqpoint{0.000000in}{0.000000in}}%
\pgfpathlineto{\pgfqpoint{0.048611in}{0.000000in}}%
\pgfusepath{stroke,fill}%
}%
\begin{pgfscope}%
\pgfsys@transformshift{5.030415in}{1.634028in}%
\pgfsys@useobject{currentmarker}{}%
\end{pgfscope}%
\end{pgfscope}%
\begin{pgfscope}%
\definecolor{textcolor}{rgb}{0.000000,0.000000,0.000000}%
\pgfsetstrokecolor{textcolor}%
\pgfsetfillcolor{textcolor}%
\pgftext[x=5.075483in,y=1.576158in,left,base]{\color{textcolor}\sffamily\fontsize{12.000000}{14.400000}\selectfont \(\displaystyle -1.5\)}%
\end{pgfscope}%
\begin{pgfscope}%
\pgfsetbuttcap%
\pgfsetroundjoin%
\definecolor{currentfill}{rgb}{0.000000,0.000000,0.000000}%
\pgfsetfillcolor{currentfill}%
\pgfsetlinewidth{0.803000pt}%
\definecolor{currentstroke}{rgb}{0.000000,0.000000,0.000000}%
\pgfsetstrokecolor{currentstroke}%
\pgfsetdash{}{0pt}%
\pgfsys@defobject{currentmarker}{\pgfqpoint{0.000000in}{0.000000in}}{\pgfqpoint{0.048611in}{0.000000in}}{%
\pgfpathmoveto{\pgfqpoint{0.000000in}{0.000000in}}%
\pgfpathlineto{\pgfqpoint{0.048611in}{0.000000in}}%
\pgfusepath{stroke,fill}%
}%
\begin{pgfscope}%
\pgfsys@transformshift{5.030415in}{2.115311in}%
\pgfsys@useobject{currentmarker}{}%
\end{pgfscope}%
\end{pgfscope}%
\begin{pgfscope}%
\definecolor{textcolor}{rgb}{0.000000,0.000000,0.000000}%
\pgfsetstrokecolor{textcolor}%
\pgfsetfillcolor{textcolor}%
\pgftext[x=5.075483in,y=2.057441in,left,base]{\color{textcolor}\sffamily\fontsize{12.000000}{14.400000}\selectfont \(\displaystyle -1.0\)}%
\end{pgfscope}%
\begin{pgfscope}%
\pgfsetbuttcap%
\pgfsetroundjoin%
\definecolor{currentfill}{rgb}{0.000000,0.000000,0.000000}%
\pgfsetfillcolor{currentfill}%
\pgfsetlinewidth{0.803000pt}%
\definecolor{currentstroke}{rgb}{0.000000,0.000000,0.000000}%
\pgfsetstrokecolor{currentstroke}%
\pgfsetdash{}{0pt}%
\pgfsys@defobject{currentmarker}{\pgfqpoint{0.000000in}{0.000000in}}{\pgfqpoint{0.048611in}{0.000000in}}{%
\pgfpathmoveto{\pgfqpoint{0.000000in}{0.000000in}}%
\pgfpathlineto{\pgfqpoint{0.048611in}{0.000000in}}%
\pgfusepath{stroke,fill}%
}%
\begin{pgfscope}%
\pgfsys@transformshift{5.030415in}{2.596594in}%
\pgfsys@useobject{currentmarker}{}%
\end{pgfscope}%
\end{pgfscope}%
\begin{pgfscope}%
\definecolor{textcolor}{rgb}{0.000000,0.000000,0.000000}%
\pgfsetstrokecolor{textcolor}%
\pgfsetfillcolor{textcolor}%
\pgftext[x=5.075483in,y=2.538724in,left,base]{\color{textcolor}\sffamily\fontsize{12.000000}{14.400000}\selectfont \(\displaystyle -0.5\)}%
\end{pgfscope}%
\begin{pgfscope}%
\pgfsetbuttcap%
\pgfsetroundjoin%
\definecolor{currentfill}{rgb}{0.000000,0.000000,0.000000}%
\pgfsetfillcolor{currentfill}%
\pgfsetlinewidth{0.803000pt}%
\definecolor{currentstroke}{rgb}{0.000000,0.000000,0.000000}%
\pgfsetstrokecolor{currentstroke}%
\pgfsetdash{}{0pt}%
\pgfsys@defobject{currentmarker}{\pgfqpoint{0.000000in}{0.000000in}}{\pgfqpoint{0.048611in}{0.000000in}}{%
\pgfpathmoveto{\pgfqpoint{0.000000in}{0.000000in}}%
\pgfpathlineto{\pgfqpoint{0.048611in}{0.000000in}}%
\pgfusepath{stroke,fill}%
}%
\begin{pgfscope}%
\pgfsys@transformshift{5.030415in}{3.077877in}%
\pgfsys@useobject{currentmarker}{}%
\end{pgfscope}%
\end{pgfscope}%
\begin{pgfscope}%
\definecolor{textcolor}{rgb}{0.000000,0.000000,0.000000}%
\pgfsetstrokecolor{textcolor}%
\pgfsetfillcolor{textcolor}%
\pgftext[x=5.205113in,y=3.020007in,left,base]{\color{textcolor}\sffamily\fontsize{12.000000}{14.400000}\selectfont \(\displaystyle 0.0\)}%
\end{pgfscope}%
\begin{pgfscope}%
\pgfsetbuttcap%
\pgfsetroundjoin%
\definecolor{currentfill}{rgb}{0.000000,0.000000,0.000000}%
\pgfsetfillcolor{currentfill}%
\pgfsetlinewidth{0.803000pt}%
\definecolor{currentstroke}{rgb}{0.000000,0.000000,0.000000}%
\pgfsetstrokecolor{currentstroke}%
\pgfsetdash{}{0pt}%
\pgfsys@defobject{currentmarker}{\pgfqpoint{0.000000in}{0.000000in}}{\pgfqpoint{0.048611in}{0.000000in}}{%
\pgfpathmoveto{\pgfqpoint{0.000000in}{0.000000in}}%
\pgfpathlineto{\pgfqpoint{0.048611in}{0.000000in}}%
\pgfusepath{stroke,fill}%
}%
\begin{pgfscope}%
\pgfsys@transformshift{5.030415in}{3.559160in}%
\pgfsys@useobject{currentmarker}{}%
\end{pgfscope}%
\end{pgfscope}%
\begin{pgfscope}%
\definecolor{textcolor}{rgb}{0.000000,0.000000,0.000000}%
\pgfsetstrokecolor{textcolor}%
\pgfsetfillcolor{textcolor}%
\pgftext[x=5.205113in,y=3.501290in,left,base]{\color{textcolor}\sffamily\fontsize{12.000000}{14.400000}\selectfont \(\displaystyle 0.5\)}%
\end{pgfscope}%
\begin{pgfscope}%
\pgfsetbuttcap%
\pgfsetroundjoin%
\definecolor{currentfill}{rgb}{0.000000,0.000000,0.000000}%
\pgfsetfillcolor{currentfill}%
\pgfsetlinewidth{0.803000pt}%
\definecolor{currentstroke}{rgb}{0.000000,0.000000,0.000000}%
\pgfsetstrokecolor{currentstroke}%
\pgfsetdash{}{0pt}%
\pgfsys@defobject{currentmarker}{\pgfqpoint{0.000000in}{0.000000in}}{\pgfqpoint{0.048611in}{0.000000in}}{%
\pgfpathmoveto{\pgfqpoint{0.000000in}{0.000000in}}%
\pgfpathlineto{\pgfqpoint{0.048611in}{0.000000in}}%
\pgfusepath{stroke,fill}%
}%
\begin{pgfscope}%
\pgfsys@transformshift{5.030415in}{4.040443in}%
\pgfsys@useobject{currentmarker}{}%
\end{pgfscope}%
\end{pgfscope}%
\begin{pgfscope}%
\definecolor{textcolor}{rgb}{0.000000,0.000000,0.000000}%
\pgfsetstrokecolor{textcolor}%
\pgfsetfillcolor{textcolor}%
\pgftext[x=5.205113in,y=3.982572in,left,base]{\color{textcolor}\sffamily\fontsize{12.000000}{14.400000}\selectfont \(\displaystyle 1.0\)}%
\end{pgfscope}%
\begin{pgfscope}%
\pgfsetbuttcap%
\pgfsetmiterjoin%
\pgfsetlinewidth{0.803000pt}%
\definecolor{currentstroke}{rgb}{0.000000,0.000000,0.000000}%
\pgfsetstrokecolor{currentstroke}%
\pgfsetdash{}{0pt}%
\pgfpathmoveto{\pgfqpoint{4.810415in}{0.345370in}}%
\pgfpathlineto{\pgfqpoint{4.810415in}{0.359807in}}%
\pgfpathlineto{\pgfqpoint{4.810415in}{4.026932in}}%
\pgfpathlineto{\pgfqpoint{4.810415in}{4.041370in}}%
\pgfpathlineto{\pgfqpoint{5.030415in}{4.041370in}}%
\pgfpathlineto{\pgfqpoint{5.030415in}{4.026932in}}%
\pgfpathlineto{\pgfqpoint{5.030415in}{0.359807in}}%
\pgfpathlineto{\pgfqpoint{5.030415in}{0.345370in}}%
\pgfpathclose%
\pgfusepath{stroke}%
\end{pgfscope}%
\end{pgfpicture}%
\makeatother%
\endgroup%

%% file: phase_s11.pgf
\begingroup%
\makeatletter%
\begin{pgfpicture}%
\pgfpathrectangle{\pgfpointorigin}{\pgfqpoint{5.537436in}{4.340290in}}%
\pgfusepath{use as bounding box, clip}%
\begin{pgfscope}%
\pgfsetbuttcap%
\pgfsetmiterjoin%
\definecolor{currentfill}{rgb}{1.000000,1.000000,1.000000}%
\pgfsetfillcolor{currentfill}%
\pgfsetlinewidth{0.000000pt}%
\definecolor{currentstroke}{rgb}{1.000000,1.000000,1.000000}%
\pgfsetstrokecolor{currentstroke}%
\pgfsetdash{}{0pt}%
\pgfpathmoveto{\pgfqpoint{0.000000in}{0.000000in}}%
\pgfpathlineto{\pgfqpoint{5.537436in}{0.000000in}}%
\pgfpathlineto{\pgfqpoint{5.537436in}{4.340290in}}%
\pgfpathlineto{\pgfqpoint{0.000000in}{4.340290in}}%
\pgfpathclose%
\pgfusepath{fill}%
\end{pgfscope}%
\begin{pgfscope}%
\pgfsetbuttcap%
\pgfsetmiterjoin%
\definecolor{currentfill}{rgb}{1.000000,1.000000,1.000000}%
\pgfsetfillcolor{currentfill}%
\pgfsetlinewidth{0.000000pt}%
\definecolor{currentstroke}{rgb}{0.000000,0.000000,0.000000}%
\pgfsetstrokecolor{currentstroke}%
\pgfsetstrokeopacity{0.000000}%
\pgfsetdash{}{0pt}%
\pgfpathmoveto{\pgfqpoint{0.360415in}{0.345370in}}%
\pgfpathlineto{\pgfqpoint{4.760415in}{0.345370in}}%
\pgfpathlineto{\pgfqpoint{4.760415in}{4.041370in}}%
\pgfpathlineto{\pgfqpoint{0.360415in}{4.041370in}}%
\pgfpathclose%
\pgfusepath{fill}%
\end{pgfscope}%
\begin{pgfscope}%
\pgfpathrectangle{\pgfqpoint{0.360415in}{0.345370in}}{\pgfqpoint{4.400000in}{3.696000in}}%
\pgfusepath{clip}%
\pgfsys@transformcm{4.400000}{0.000000}{0.000000}{-3.700000}{0.360415in}{4.045370in}%
\pgftext[left,bottom]{\pgfimage[interpolate=false,width=1.000000in,height=1.000000in]{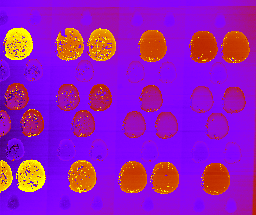}}%
\end{pgfscope}%
\begin{pgfscope}%
\pgfsetbuttcap%
\pgfsetroundjoin%
\definecolor{currentfill}{rgb}{0.000000,0.000000,0.000000}%
\pgfsetfillcolor{currentfill}%
\pgfsetlinewidth{0.803000pt}%
\definecolor{currentstroke}{rgb}{0.000000,0.000000,0.000000}%
\pgfsetstrokecolor{currentstroke}%
\pgfsetdash{}{0pt}%
\pgfsys@defobject{currentmarker}{\pgfqpoint{0.000000in}{-0.048611in}}{\pgfqpoint{0.000000in}{0.000000in}}{%
\pgfpathmoveto{\pgfqpoint{0.000000in}{0.000000in}}%
\pgfpathlineto{\pgfqpoint{0.000000in}{-0.048611in}}%
\pgfusepath{stroke,fill}%
}%
\begin{pgfscope}%
\pgfsys@transformshift{0.360415in}{0.345370in}%
\pgfsys@useobject{currentmarker}{}%
\end{pgfscope}%
\end{pgfscope}%
\begin{pgfscope}%
\definecolor{textcolor}{rgb}{0.000000,0.000000,0.000000}%
\pgfsetstrokecolor{textcolor}%
\pgfsetfillcolor{textcolor}%
\pgftext[x=0.360415in,y=0.248148in,,top]{\color{textcolor}\sffamily\fontsize{12.000000}{14.400000}\selectfont \(\displaystyle 0\)}%
\end{pgfscope}%
\begin{pgfscope}%
\pgfsetbuttcap%
\pgfsetroundjoin%
\definecolor{currentfill}{rgb}{0.000000,0.000000,0.000000}%
\pgfsetfillcolor{currentfill}%
\pgfsetlinewidth{0.803000pt}%
\definecolor{currentstroke}{rgb}{0.000000,0.000000,0.000000}%
\pgfsetstrokecolor{currentstroke}%
\pgfsetdash{}{0pt}%
\pgfsys@defobject{currentmarker}{\pgfqpoint{0.000000in}{-0.048611in}}{\pgfqpoint{0.000000in}{0.000000in}}{%
\pgfpathmoveto{\pgfqpoint{0.000000in}{0.000000in}}%
\pgfpathlineto{\pgfqpoint{0.000000in}{-0.048611in}}%
\pgfusepath{stroke,fill}%
}%
\begin{pgfscope}%
\pgfsys@transformshift{1.240415in}{0.345370in}%
\pgfsys@useobject{currentmarker}{}%
\end{pgfscope}%
\end{pgfscope}%
\begin{pgfscope}%
\definecolor{textcolor}{rgb}{0.000000,0.000000,0.000000}%
\pgfsetstrokecolor{textcolor}%
\pgfsetfillcolor{textcolor}%
\pgftext[x=1.240415in,y=0.248148in,,top]{\color{textcolor}\sffamily\fontsize{12.000000}{14.400000}\selectfont \(\displaystyle 10\)}%
\end{pgfscope}%
\begin{pgfscope}%
\pgfsetbuttcap%
\pgfsetroundjoin%
\definecolor{currentfill}{rgb}{0.000000,0.000000,0.000000}%
\pgfsetfillcolor{currentfill}%
\pgfsetlinewidth{0.803000pt}%
\definecolor{currentstroke}{rgb}{0.000000,0.000000,0.000000}%
\pgfsetstrokecolor{currentstroke}%
\pgfsetdash{}{0pt}%
\pgfsys@defobject{currentmarker}{\pgfqpoint{0.000000in}{-0.048611in}}{\pgfqpoint{0.000000in}{0.000000in}}{%
\pgfpathmoveto{\pgfqpoint{0.000000in}{0.000000in}}%
\pgfpathlineto{\pgfqpoint{0.000000in}{-0.048611in}}%
\pgfusepath{stroke,fill}%
}%
\begin{pgfscope}%
\pgfsys@transformshift{2.120415in}{0.345370in}%
\pgfsys@useobject{currentmarker}{}%
\end{pgfscope}%
\end{pgfscope}%
\begin{pgfscope}%
\definecolor{textcolor}{rgb}{0.000000,0.000000,0.000000}%
\pgfsetstrokecolor{textcolor}%
\pgfsetfillcolor{textcolor}%
\pgftext[x=2.120415in,y=0.248148in,,top]{\color{textcolor}\sffamily\fontsize{12.000000}{14.400000}\selectfont \(\displaystyle 20\)}%
\end{pgfscope}%
\begin{pgfscope}%
\pgfsetbuttcap%
\pgfsetroundjoin%
\definecolor{currentfill}{rgb}{0.000000,0.000000,0.000000}%
\pgfsetfillcolor{currentfill}%
\pgfsetlinewidth{0.803000pt}%
\definecolor{currentstroke}{rgb}{0.000000,0.000000,0.000000}%
\pgfsetstrokecolor{currentstroke}%
\pgfsetdash{}{0pt}%
\pgfsys@defobject{currentmarker}{\pgfqpoint{0.000000in}{-0.048611in}}{\pgfqpoint{0.000000in}{0.000000in}}{%
\pgfpathmoveto{\pgfqpoint{0.000000in}{0.000000in}}%
\pgfpathlineto{\pgfqpoint{0.000000in}{-0.048611in}}%
\pgfusepath{stroke,fill}%
}%
\begin{pgfscope}%
\pgfsys@transformshift{3.000415in}{0.345370in}%
\pgfsys@useobject{currentmarker}{}%
\end{pgfscope}%
\end{pgfscope}%
\begin{pgfscope}%
\definecolor{textcolor}{rgb}{0.000000,0.000000,0.000000}%
\pgfsetstrokecolor{textcolor}%
\pgfsetfillcolor{textcolor}%
\pgftext[x=3.000415in,y=0.248148in,,top]{\color{textcolor}\sffamily\fontsize{12.000000}{14.400000}\selectfont \(\displaystyle 30\)}%
\end{pgfscope}%
\begin{pgfscope}%
\pgfsetbuttcap%
\pgfsetroundjoin%
\definecolor{currentfill}{rgb}{0.000000,0.000000,0.000000}%
\pgfsetfillcolor{currentfill}%
\pgfsetlinewidth{0.803000pt}%
\definecolor{currentstroke}{rgb}{0.000000,0.000000,0.000000}%
\pgfsetstrokecolor{currentstroke}%
\pgfsetdash{}{0pt}%
\pgfsys@defobject{currentmarker}{\pgfqpoint{0.000000in}{-0.048611in}}{\pgfqpoint{0.000000in}{0.000000in}}{%
\pgfpathmoveto{\pgfqpoint{0.000000in}{0.000000in}}%
\pgfpathlineto{\pgfqpoint{0.000000in}{-0.048611in}}%
\pgfusepath{stroke,fill}%
}%
\begin{pgfscope}%
\pgfsys@transformshift{3.880415in}{0.345370in}%
\pgfsys@useobject{currentmarker}{}%
\end{pgfscope}%
\end{pgfscope}%
\begin{pgfscope}%
\definecolor{textcolor}{rgb}{0.000000,0.000000,0.000000}%
\pgfsetstrokecolor{textcolor}%
\pgfsetfillcolor{textcolor}%
\pgftext[x=3.880415in,y=0.248148in,,top]{\color{textcolor}\sffamily\fontsize{12.000000}{14.400000}\selectfont \(\displaystyle 40\)}%
\end{pgfscope}%
\begin{pgfscope}%
\pgfsetbuttcap%
\pgfsetroundjoin%
\definecolor{currentfill}{rgb}{0.000000,0.000000,0.000000}%
\pgfsetfillcolor{currentfill}%
\pgfsetlinewidth{0.803000pt}%
\definecolor{currentstroke}{rgb}{0.000000,0.000000,0.000000}%
\pgfsetstrokecolor{currentstroke}%
\pgfsetdash{}{0pt}%
\pgfsys@defobject{currentmarker}{\pgfqpoint{0.000000in}{-0.048611in}}{\pgfqpoint{0.000000in}{0.000000in}}{%
\pgfpathmoveto{\pgfqpoint{0.000000in}{0.000000in}}%
\pgfpathlineto{\pgfqpoint{0.000000in}{-0.048611in}}%
\pgfusepath{stroke,fill}%
}%
\begin{pgfscope}%
\pgfsys@transformshift{4.760415in}{0.345370in}%
\pgfsys@useobject{currentmarker}{}%
\end{pgfscope}%
\end{pgfscope}%
\begin{pgfscope}%
\definecolor{textcolor}{rgb}{0.000000,0.000000,0.000000}%
\pgfsetstrokecolor{textcolor}%
\pgfsetfillcolor{textcolor}%
\pgftext[x=4.760415in,y=0.248148in,,top]{\color{textcolor}\sffamily\fontsize{12.000000}{14.400000}\selectfont \(\displaystyle 50\)}%
\end{pgfscope}%
\begin{pgfscope}%
\pgfsetbuttcap%
\pgfsetroundjoin%
\definecolor{currentfill}{rgb}{0.000000,0.000000,0.000000}%
\pgfsetfillcolor{currentfill}%
\pgfsetlinewidth{0.803000pt}%
\definecolor{currentstroke}{rgb}{0.000000,0.000000,0.000000}%
\pgfsetstrokecolor{currentstroke}%
\pgfsetdash{}{0pt}%
\pgfsys@defobject{currentmarker}{\pgfqpoint{-0.048611in}{0.000000in}}{\pgfqpoint{0.000000in}{0.000000in}}{%
\pgfpathmoveto{\pgfqpoint{0.000000in}{0.000000in}}%
\pgfpathlineto{\pgfqpoint{-0.048611in}{0.000000in}}%
\pgfusepath{stroke,fill}%
}%
\begin{pgfscope}%
\pgfsys@transformshift{0.360415in}{0.345370in}%
\pgfsys@useobject{currentmarker}{}%
\end{pgfscope}%
\end{pgfscope}%
\begin{pgfscope}%
\definecolor{textcolor}{rgb}{0.000000,0.000000,0.000000}%
\pgfsetstrokecolor{textcolor}%
\pgfsetfillcolor{textcolor}%
\pgftext[x=0.181596in,y=0.287500in,left,base]{\color{textcolor}\sffamily\fontsize{12.000000}{14.400000}\selectfont \(\displaystyle 0\)}%
\end{pgfscope}%
\begin{pgfscope}%
\pgfsetbuttcap%
\pgfsetroundjoin%
\definecolor{currentfill}{rgb}{0.000000,0.000000,0.000000}%
\pgfsetfillcolor{currentfill}%
\pgfsetlinewidth{0.803000pt}%
\definecolor{currentstroke}{rgb}{0.000000,0.000000,0.000000}%
\pgfsetstrokecolor{currentstroke}%
\pgfsetdash{}{0pt}%
\pgfsys@defobject{currentmarker}{\pgfqpoint{-0.048611in}{0.000000in}}{\pgfqpoint{0.000000in}{0.000000in}}{%
\pgfpathmoveto{\pgfqpoint{0.000000in}{0.000000in}}%
\pgfpathlineto{\pgfqpoint{-0.048611in}{0.000000in}}%
\pgfusepath{stroke,fill}%
}%
\begin{pgfscope}%
\pgfsys@transformshift{0.360415in}{1.225370in}%
\pgfsys@useobject{currentmarker}{}%
\end{pgfscope}%
\end{pgfscope}%
\begin{pgfscope}%
\definecolor{textcolor}{rgb}{0.000000,0.000000,0.000000}%
\pgfsetstrokecolor{textcolor}%
\pgfsetfillcolor{textcolor}%
\pgftext[x=0.100000in,y=1.167500in,left,base]{\color{textcolor}\sffamily\fontsize{12.000000}{14.400000}\selectfont \(\displaystyle 10\)}%
\end{pgfscope}%
\begin{pgfscope}%
\pgfsetbuttcap%
\pgfsetroundjoin%
\definecolor{currentfill}{rgb}{0.000000,0.000000,0.000000}%
\pgfsetfillcolor{currentfill}%
\pgfsetlinewidth{0.803000pt}%
\definecolor{currentstroke}{rgb}{0.000000,0.000000,0.000000}%
\pgfsetstrokecolor{currentstroke}%
\pgfsetdash{}{0pt}%
\pgfsys@defobject{currentmarker}{\pgfqpoint{-0.048611in}{0.000000in}}{\pgfqpoint{0.000000in}{0.000000in}}{%
\pgfpathmoveto{\pgfqpoint{0.000000in}{0.000000in}}%
\pgfpathlineto{\pgfqpoint{-0.048611in}{0.000000in}}%
\pgfusepath{stroke,fill}%
}%
\begin{pgfscope}%
\pgfsys@transformshift{0.360415in}{2.105370in}%
\pgfsys@useobject{currentmarker}{}%
\end{pgfscope}%
\end{pgfscope}%
\begin{pgfscope}%
\definecolor{textcolor}{rgb}{0.000000,0.000000,0.000000}%
\pgfsetstrokecolor{textcolor}%
\pgfsetfillcolor{textcolor}%
\pgftext[x=0.100000in,y=2.047500in,left,base]{\color{textcolor}\sffamily\fontsize{12.000000}{14.400000}\selectfont \(\displaystyle 20\)}%
\end{pgfscope}%
\begin{pgfscope}%
\pgfsetbuttcap%
\pgfsetroundjoin%
\definecolor{currentfill}{rgb}{0.000000,0.000000,0.000000}%
\pgfsetfillcolor{currentfill}%
\pgfsetlinewidth{0.803000pt}%
\definecolor{currentstroke}{rgb}{0.000000,0.000000,0.000000}%
\pgfsetstrokecolor{currentstroke}%
\pgfsetdash{}{0pt}%
\pgfsys@defobject{currentmarker}{\pgfqpoint{-0.048611in}{0.000000in}}{\pgfqpoint{0.000000in}{0.000000in}}{%
\pgfpathmoveto{\pgfqpoint{0.000000in}{0.000000in}}%
\pgfpathlineto{\pgfqpoint{-0.048611in}{0.000000in}}%
\pgfusepath{stroke,fill}%
}%
\begin{pgfscope}%
\pgfsys@transformshift{0.360415in}{2.985370in}%
\pgfsys@useobject{currentmarker}{}%
\end{pgfscope}%
\end{pgfscope}%
\begin{pgfscope}%
\definecolor{textcolor}{rgb}{0.000000,0.000000,0.000000}%
\pgfsetstrokecolor{textcolor}%
\pgfsetfillcolor{textcolor}%
\pgftext[x=0.100000in,y=2.927500in,left,base]{\color{textcolor}\sffamily\fontsize{12.000000}{14.400000}\selectfont \(\displaystyle 30\)}%
\end{pgfscope}%
\begin{pgfscope}%
\pgfsetbuttcap%
\pgfsetroundjoin%
\definecolor{currentfill}{rgb}{0.000000,0.000000,0.000000}%
\pgfsetfillcolor{currentfill}%
\pgfsetlinewidth{0.803000pt}%
\definecolor{currentstroke}{rgb}{0.000000,0.000000,0.000000}%
\pgfsetstrokecolor{currentstroke}%
\pgfsetdash{}{0pt}%
\pgfsys@defobject{currentmarker}{\pgfqpoint{-0.048611in}{0.000000in}}{\pgfqpoint{0.000000in}{0.000000in}}{%
\pgfpathmoveto{\pgfqpoint{0.000000in}{0.000000in}}%
\pgfpathlineto{\pgfqpoint{-0.048611in}{0.000000in}}%
\pgfusepath{stroke,fill}%
}%
\begin{pgfscope}%
\pgfsys@transformshift{0.360415in}{3.865370in}%
\pgfsys@useobject{currentmarker}{}%
\end{pgfscope}%
\end{pgfscope}%
\begin{pgfscope}%
\definecolor{textcolor}{rgb}{0.000000,0.000000,0.000000}%
\pgfsetstrokecolor{textcolor}%
\pgfsetfillcolor{textcolor}%
\pgftext[x=0.100000in,y=3.807500in,left,base]{\color{textcolor}\sffamily\fontsize{12.000000}{14.400000}\selectfont \(\displaystyle 40\)}%
\end{pgfscope}%
\begin{pgfscope}%
\pgfsetrectcap%
\pgfsetmiterjoin%
\pgfsetlinewidth{0.803000pt}%
\definecolor{currentstroke}{rgb}{0.000000,0.000000,0.000000}%
\pgfsetstrokecolor{currentstroke}%
\pgfsetdash{}{0pt}%
\pgfpathmoveto{\pgfqpoint{0.360415in}{0.345370in}}%
\pgfpathlineto{\pgfqpoint{0.360415in}{4.041370in}}%
\pgfusepath{stroke}%
\end{pgfscope}%
\begin{pgfscope}%
\pgfsetrectcap%
\pgfsetmiterjoin%
\pgfsetlinewidth{0.803000pt}%
\definecolor{currentstroke}{rgb}{0.000000,0.000000,0.000000}%
\pgfsetstrokecolor{currentstroke}%
\pgfsetdash{}{0pt}%
\pgfpathmoveto{\pgfqpoint{4.760415in}{0.345370in}}%
\pgfpathlineto{\pgfqpoint{4.760415in}{4.041370in}}%
\pgfusepath{stroke}%
\end{pgfscope}%
\begin{pgfscope}%
\pgfsetrectcap%
\pgfsetmiterjoin%
\pgfsetlinewidth{0.803000pt}%
\definecolor{currentstroke}{rgb}{0.000000,0.000000,0.000000}%
\pgfsetstrokecolor{currentstroke}%
\pgfsetdash{}{0pt}%
\pgfpathmoveto{\pgfqpoint{0.360415in}{0.345370in}}%
\pgfpathlineto{\pgfqpoint{4.760415in}{0.345370in}}%
\pgfusepath{stroke}%
\end{pgfscope}%
\begin{pgfscope}%
\pgfsetrectcap%
\pgfsetmiterjoin%
\pgfsetlinewidth{0.803000pt}%
\definecolor{currentstroke}{rgb}{0.000000,0.000000,0.000000}%
\pgfsetstrokecolor{currentstroke}%
\pgfsetdash{}{0pt}%
\pgfpathmoveto{\pgfqpoint{0.360415in}{4.041370in}}%
\pgfpathlineto{\pgfqpoint{4.760415in}{4.041370in}}%
\pgfusepath{stroke}%
\end{pgfscope}%
\begin{pgfscope}%
\definecolor{textcolor}{rgb}{0.000000,0.000000,0.000000}%
\pgfsetstrokecolor{textcolor}%
\pgfsetfillcolor{textcolor}%
\pgftext[x=4.364415in,y=4.115290in,left,base]{\color{textcolor}\sffamily\fontsize{12.000000}{14.400000}\selectfont Arg(\(\displaystyle S11\))/mrad}%
\end{pgfscope}%
\begin{pgfscope}%
\definecolor{textcolor}{rgb}{0.000000,0.000000,0.000000}%
\pgfsetstrokecolor{textcolor}%
\pgfsetfillcolor{textcolor}%
\pgftext[x=0.897215in,y=3.372570in,,]{\color{textcolor}\sffamily\fontsize{18.000000}{21.600000}\selectfont \(\displaystyle \mathbf{C_1}\)}%
\end{pgfscope}%
\begin{pgfscope}%
\definecolor{textcolor}{rgb}{0.000000,0.000000,0.000000}%
\pgfsetstrokecolor{textcolor}%
\pgfsetfillcolor{textcolor}%
\pgftext[x=4.338015in,y=2.985370in,,]{\color{textcolor}\sffamily\fontsize{18.000000}{21.600000}\selectfont \(\displaystyle \mathbf{C_2}\)}%
\end{pgfscope}%
\begin{pgfscope}%
\definecolor{textcolor}{rgb}{0.000000,0.000000,0.000000}%
\pgfsetstrokecolor{textcolor}%
\pgfsetfillcolor{textcolor}%
\pgftext[x=0.448415in,y=0.142090in,left,base]{\color{textcolor}\sffamily\fontsize{12.000000}{14.400000}\selectfont \(\displaystyle \mu m\)}%
\end{pgfscope}%
\begin{pgfscope}%
\pgfpathrectangle{\pgfqpoint{4.810415in}{0.345370in}}{\pgfqpoint{0.220000in}{3.696000in}}%
\pgfusepath{clip}%
\pgfsetbuttcap%
\pgfsetmiterjoin%
\definecolor{currentfill}{rgb}{1.000000,1.000000,1.000000}%
\pgfsetfillcolor{currentfill}%
\pgfsetlinewidth{0.010037pt}%
\definecolor{currentstroke}{rgb}{1.000000,1.000000,1.000000}%
\pgfsetstrokecolor{currentstroke}%
\pgfsetdash{}{0pt}%
\pgfpathmoveto{\pgfqpoint{4.810415in}{0.345370in}}%
\pgfpathlineto{\pgfqpoint{4.810415in}{0.359807in}}%
\pgfpathlineto{\pgfqpoint{4.810415in}{4.026932in}}%
\pgfpathlineto{\pgfqpoint{4.810415in}{4.041370in}}%
\pgfpathlineto{\pgfqpoint{5.030415in}{4.041370in}}%
\pgfpathlineto{\pgfqpoint{5.030415in}{4.026932in}}%
\pgfpathlineto{\pgfqpoint{5.030415in}{0.359807in}}%
\pgfpathlineto{\pgfqpoint{5.030415in}{0.345370in}}%
\pgfpathclose%
\pgfusepath{stroke,fill}%
\end{pgfscope}%
\begin{pgfscope}%
\pgfsys@transformshift{4.810000in}{0.350290in}%
\pgftext[left,bottom]{\pgfimage[interpolate=true,width=0.220000in,height=3.690000in]{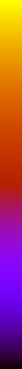}}%
\end{pgfscope}%
\begin{pgfscope}%
\pgfsetbuttcap%
\pgfsetroundjoin%
\definecolor{currentfill}{rgb}{0.000000,0.000000,0.000000}%
\pgfsetfillcolor{currentfill}%
\pgfsetlinewidth{0.803000pt}%
\definecolor{currentstroke}{rgb}{0.000000,0.000000,0.000000}%
\pgfsetstrokecolor{currentstroke}%
\pgfsetdash{}{0pt}%
\pgfsys@defobject{currentmarker}{\pgfqpoint{0.000000in}{0.000000in}}{\pgfqpoint{0.048611in}{0.000000in}}{%
\pgfpathmoveto{\pgfqpoint{0.000000in}{0.000000in}}%
\pgfpathlineto{\pgfqpoint{0.048611in}{0.000000in}}%
\pgfusepath{stroke,fill}%
}%
\begin{pgfscope}%
\pgfsys@transformshift{5.030415in}{0.895567in}%
\pgfsys@useobject{currentmarker}{}%
\end{pgfscope}%
\end{pgfscope}%
\begin{pgfscope}%
\definecolor{textcolor}{rgb}{0.000000,0.000000,0.000000}%
\pgfsetstrokecolor{textcolor}%
\pgfsetfillcolor{textcolor}%
\pgftext[x=5.202411in,y=0.837697in,left,base]{\color{textcolor}\sffamily\fontsize{12.000000}{14.400000}\selectfont \(\displaystyle -2\)}%
\end{pgfscope}%
\begin{pgfscope}%
\pgfsetbuttcap%
\pgfsetroundjoin%
\definecolor{currentfill}{rgb}{0.000000,0.000000,0.000000}%
\pgfsetfillcolor{currentfill}%
\pgfsetlinewidth{0.803000pt}%
\definecolor{currentstroke}{rgb}{0.000000,0.000000,0.000000}%
\pgfsetstrokecolor{currentstroke}%
\pgfsetdash{}{0pt}%
\pgfsys@defobject{currentmarker}{\pgfqpoint{0.000000in}{0.000000in}}{\pgfqpoint{0.048611in}{0.000000in}}{%
\pgfpathmoveto{\pgfqpoint{0.000000in}{0.000000in}}%
\pgfpathlineto{\pgfqpoint{0.048611in}{0.000000in}}%
\pgfusepath{stroke,fill}%
}%
\begin{pgfscope}%
\pgfsys@transformshift{5.030415in}{1.682018in}%
\pgfsys@useobject{currentmarker}{}%
\end{pgfscope}%
\end{pgfscope}%
\begin{pgfscope}%
\definecolor{textcolor}{rgb}{0.000000,0.000000,0.000000}%
\pgfsetstrokecolor{textcolor}%
\pgfsetfillcolor{textcolor}%
\pgftext[x=5.332041in,y=1.624147in,left,base]{\color{textcolor}\sffamily\fontsize{12.000000}{14.400000}\selectfont \(\displaystyle 0\)}%
\end{pgfscope}%
\begin{pgfscope}%
\pgfsetbuttcap%
\pgfsetroundjoin%
\definecolor{currentfill}{rgb}{0.000000,0.000000,0.000000}%
\pgfsetfillcolor{currentfill}%
\pgfsetlinewidth{0.803000pt}%
\definecolor{currentstroke}{rgb}{0.000000,0.000000,0.000000}%
\pgfsetstrokecolor{currentstroke}%
\pgfsetdash{}{0pt}%
\pgfsys@defobject{currentmarker}{\pgfqpoint{0.000000in}{0.000000in}}{\pgfqpoint{0.048611in}{0.000000in}}{%
\pgfpathmoveto{\pgfqpoint{0.000000in}{0.000000in}}%
\pgfpathlineto{\pgfqpoint{0.048611in}{0.000000in}}%
\pgfusepath{stroke,fill}%
}%
\begin{pgfscope}%
\pgfsys@transformshift{5.030415in}{2.468468in}%
\pgfsys@useobject{currentmarker}{}%
\end{pgfscope}%
\end{pgfscope}%
\begin{pgfscope}%
\definecolor{textcolor}{rgb}{0.000000,0.000000,0.000000}%
\pgfsetstrokecolor{textcolor}%
\pgfsetfillcolor{textcolor}%
\pgftext[x=5.332041in,y=2.410598in,left,base]{\color{textcolor}\sffamily\fontsize{12.000000}{14.400000}\selectfont \(\displaystyle 2\)}%
\end{pgfscope}%
\begin{pgfscope}%
\pgfsetbuttcap%
\pgfsetroundjoin%
\definecolor{currentfill}{rgb}{0.000000,0.000000,0.000000}%
\pgfsetfillcolor{currentfill}%
\pgfsetlinewidth{0.803000pt}%
\definecolor{currentstroke}{rgb}{0.000000,0.000000,0.000000}%
\pgfsetstrokecolor{currentstroke}%
\pgfsetdash{}{0pt}%
\pgfsys@defobject{currentmarker}{\pgfqpoint{0.000000in}{0.000000in}}{\pgfqpoint{0.048611in}{0.000000in}}{%
\pgfpathmoveto{\pgfqpoint{0.000000in}{0.000000in}}%
\pgfpathlineto{\pgfqpoint{0.048611in}{0.000000in}}%
\pgfusepath{stroke,fill}%
}%
\begin{pgfscope}%
\pgfsys@transformshift{5.030415in}{3.254919in}%
\pgfsys@useobject{currentmarker}{}%
\end{pgfscope}%
\end{pgfscope}%
\begin{pgfscope}%
\definecolor{textcolor}{rgb}{0.000000,0.000000,0.000000}%
\pgfsetstrokecolor{textcolor}%
\pgfsetfillcolor{textcolor}%
\pgftext[x=5.332041in,y=3.197049in,left,base]{\color{textcolor}\sffamily\fontsize{12.000000}{14.400000}\selectfont \(\displaystyle 4\)}%
\end{pgfscope}%
\begin{pgfscope}%
\pgfsetbuttcap%
\pgfsetroundjoin%
\definecolor{currentfill}{rgb}{0.000000,0.000000,0.000000}%
\pgfsetfillcolor{currentfill}%
\pgfsetlinewidth{0.803000pt}%
\definecolor{currentstroke}{rgb}{0.000000,0.000000,0.000000}%
\pgfsetstrokecolor{currentstroke}%
\pgfsetdash{}{0pt}%
\pgfsys@defobject{currentmarker}{\pgfqpoint{0.000000in}{0.000000in}}{\pgfqpoint{0.048611in}{0.000000in}}{%
\pgfpathmoveto{\pgfqpoint{0.000000in}{0.000000in}}%
\pgfpathlineto{\pgfqpoint{0.048611in}{0.000000in}}%
\pgfusepath{stroke,fill}%
}%
\begin{pgfscope}%
\pgfsys@transformshift{5.030415in}{4.041370in}%
\pgfsys@useobject{currentmarker}{}%
\end{pgfscope}%
\end{pgfscope}%
\begin{pgfscope}%
\definecolor{textcolor}{rgb}{0.000000,0.000000,0.000000}%
\pgfsetstrokecolor{textcolor}%
\pgfsetfillcolor{textcolor}%
\pgftext[x=5.332041in,y=3.983500in,left,base]{\color{textcolor}\sffamily\fontsize{12.000000}{14.400000}\selectfont \(\displaystyle 6\)}%
\end{pgfscope}%
\begin{pgfscope}%
\pgfsetbuttcap%
\pgfsetmiterjoin%
\pgfsetlinewidth{0.803000pt}%
\definecolor{currentstroke}{rgb}{0.000000,0.000000,0.000000}%
\pgfsetstrokecolor{currentstroke}%
\pgfsetdash{}{0pt}%
\pgfpathmoveto{\pgfqpoint{4.810415in}{0.345370in}}%
\pgfpathlineto{\pgfqpoint{4.810415in}{0.359807in}}%
\pgfpathlineto{\pgfqpoint{4.810415in}{4.026932in}}%
\pgfpathlineto{\pgfqpoint{4.810415in}{4.041370in}}%
\pgfpathlineto{\pgfqpoint{5.030415in}{4.041370in}}%
\pgfpathlineto{\pgfqpoint{5.030415in}{4.026932in}}%
\pgfpathlineto{\pgfqpoint{5.030415in}{0.359807in}}%
\pgfpathlineto{\pgfqpoint{5.030415in}{0.345370in}}%
\pgfpathclose%
\pgfusepath{stroke}%
\end{pgfscope}%
\end{pgfpicture}%
\makeatother%
\endgroup%